\documentclass[12pt,showpacs,aps,prd,nofootinbib,floatfix,amsmath,amssymb]{revtex4}
\usepackage{graphicx}
\usepackage[usenames]{color}

\begin{document}

\title{Sensitivity of the decay $h\to ZZ^*\to Zl+l-$ to the Higgs self coupling
    through radiative corrections}

\author{H. Castilla-Valdez}
\email{castilla@fis.cinvestav.mx}
\affiliation{Departamento de F\'{\i}sica, CINVESTAV, Apartado
Postal 14-740, 07000 M\'exico, D. F., M\'exico.}
\author{C. G. Honorato }
\email{carlosg.honorato@correo.buap.mx}
\affiliation{Fac. de Cs. de la Electr\'onica, Benem\'erita Universidad Aut\'onoma de Puebla, Apartado Postal 542, 72570 Puebla, Puebla, M\'exico}
\author{A. Moyotl}
\email{amoyotl@fis.cinvestav.mx}
\affiliation{Departamento de F\'{\i}sica, CINVESTAV, Apartado
Postal 14-740, 07000 M\'exico, D. F., M\'exico.}
\author{M. A. Perez }
\email{mperez@fis.cinvestav.mx}
\affiliation{Departamento de F\'{\i}sica, CINVESTAV, Apartado
Postal 14-740, 07000 M\'exico, D. F., M\'exico.}


\date{\today}

\begin{abstract}
      We study the radiative corrections induced by the triple Higgs
      boson coupling $hhh$ in the three body decay $h\to ZZ^*\to Zl\bar l$. We
      show that these corrections are potentially sensitive to the
      specific value of this coupling in the Standard Model and the
      Two Higgs Doublet Model (2HDM). These effects may induce
      corrections to the integrated decay width of the three-body
      decay  of order few percent in the 2HDM and thus open a new window to test
      the Higgs boson self interaction in physics beyond the standard model.
.
\end{abstract}
\maketitle

\section{Introduction}

 The accumulated data at the LHC has confirmed that the new particle with a mass of about
 $125$ $GeV$ corresponds to the state responsible for the
 electroweak symmetry breaking  mechanism of the Standard
 Model (SM) \cite{Atlas,PHiggs}. This data indicates also that the
 couplings of this state to fermions and gauge bosons are
 consistent with those expected in the SM for the Higgs
 boson \cite{Aad}. An immediate task now is to measure the Higgs
 self-coupling $hhh$ which will determine the structure of the
 Higgs potential in the SM. Measuring this coupling will
 be then an important step to conclude that the observed
 scalar boson \cite{Atlas,PHiggs} is identical to the Higgs boson
 predicted by the SM.

 It has been pointed out that the $hhh$ coupling may be accessible
 in the double Higgs production in both $e^+e^-$ linear colliders
 \cite{Aguilar,Gutierrez}  and at the LHC \cite{Baur, Djouadi}.
 However, the production cross section in the latter case is about two orders of magnitude
 below the single Higgs production case \cite{Dolan}. The gluon-gluon
(GGF) and vector boson fusion (VBF) modes seem to be the most
 sensitive channels to the $hhh$ contribution for the double Higgs
 production process \cite{Baglio, Dawson, Maltoni}. Unfortunately, the respective cross
 sections for these modes have to be measured with an accuracy
 of about $50\%$ at $\sqrt{s}=8$ $TeV$ in order to be able to extract the
 trilinear coupling with a similar accuracy \cite{Baglio} . The situation
 may be improved at a $100$ $TeV$ hadron collider with a $bb\gamma\gamma$ final
 state  \cite{He}. On the other hand, it has been pointed out that
 a precise measurement of the ratio of cross sections of the
 double-to-single Higgs boson production at the LHC may become
 the most precise method for determination of the Higgs trilinear
 coupling \cite{Goertz}. However, it is not clear  if a
 meaningful measurement of the Higgs self-coupling at the
 LHC is possible due to the relative high uncertainties
 on the HH cross sections measurements in both CMS and ATLAS \cite{Flechl}.

It has been known that indirect tests of new physics effects can
be performed by precision measurements of observables sensitive to
radiative corrections \cite{Martinez}. In particular, the study of deviations of
the Higgs boson couplings from the SM predictions may discriminate
among various new physics models.  The measurement accuracy of
these coupling constants will be improved at future experiments
such as the High Luminosty LHC (HL-LHC), and even most of the
Higgs couplings are expected to be measured with typically of
 $10\%$ or better accuracy \cite{Kanemura1, Heinemeyer, Cao}. In the present paper we are
interested in testing if the trilinear self-coupling of the Higgs
boson may be detected through its virtual effects in the radiative
corrections to  the decay mode $h\to ZZ^*\to Zl^+l^-$. We will
obtain that these corrections are potentially sensitive to the
specific value of the Higgs self-coupling in the SM and in the
2HDM. The
      one-loop effects for the $hZZ$ coupling has been computed in
      the SM \cite{Cao}, the 2HDM \cite{Kanemura1} and the Inert Higgs Doublet
      Model (IHDM) \cite{Arhrib}. In all these cases, the radiative
      corrections are small between $1\%$ and $2\%$. However, we will
      show that variations on the $hhh$ coupling may induce higher
      corrections on the partial decay width of $h\to ZZ^*\to \to
      Zl^+l^-$   and thus open a window to test the self Higgs coupling
      if the respective decay width can be measured with an
      accuracy of order $4\%$. Deviations of the triple self coupling with respect to the
    SM predictions has been observed in the effective Lagrangian
    framework 2HDM  \cite{Kanemura1} and in radiative corrections in 2HDM
     \cite{Castilla,Figy}.

The plan of the paper is as follows. In Section II we
    present the details of the radiative corrections to one-loop order of the decay width for $h\to ZZ^*\to Zl^+l^-$ in the SM;
    section III contains the respective calculation for the
    the 2HDM. Conclusions are given in Section IV while the
    Appendix includes the definitions for the different
    structure funcions used in our calculation.

\section{SM FRAMEWORK}

This decay process occurs, at first glance, in the SM context. Actually, it is achieved at tree level, but the relevant effects due the $hhh$ vertex, are induced at one loop level. With the general aim to present a clear analysis over de Higgs potential we will use the following notation 
\begin{equation}
  V_h=\mu^2\Phi\Phi^\dagger+\frac{1}{2}
  \lambda(\Phi\Phi^\dagger)^2,
\end{equation}
with $\Phi$ the Higgs doublet and after the EWSB with $v$ the
vacuum expectation value the Higgs potential takes the expression
\begin{equation}
V_h= \frac{1}{2}m_h^2 h^2 + \lambda_{3h}v(h)^3 +
\frac{1}{4}\lambda_{4h}h^4,
\end{equation}
with $\lambda_{3h}=\frac{m_h^2}{2v^2}$ and $\lambda_{4h}
=\lambda_{3h}$ in the SM. For a Higgs mass of  $125$ $GeV$ and
$v= 248$ $GeV$ we get $\lambda_{3h}=0.13$. We will find
convenient to use a normalized Higgs self coupling \cite{Dolan}:
  \begin{equation}\label{lambda}
   \lambda = \frac{\lambda_{NP}}{\lambda_{SM}}.
  \end{equation}


\begin{figure}[h]
\begin{center}
\includegraphics[width=2 in]{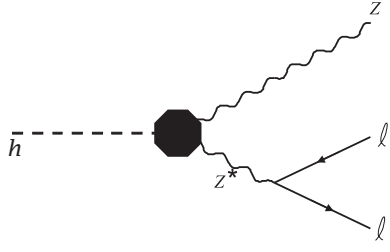}\\
\caption{Feynman diagram for decay $h\to ZZ^*\to
Zll$.} \label{fig-decay}
\end{center}
\end{figure}

    The Higgs decay $h\to ZZ^* \to Zl^+l^-$ proceeds through the Feynman
    diagram shown in Figure \ref{fig-decay}. The black vertex represents all
    the perturbative contributions that are included in Figures
    2-5 to one-loop order. The integrated decay width is given
    by  \cite{Bager}:
\begin{eqnarray} \label{Decay-SM}
\Gamma(h_{SM}\to Zll)&=&\frac{m_{h}}{256\pi^3}\int_{x_{1i}}^{x_{1f}}\int_{x_{2i}}^{x_{2f}}|\bar{\cal M}|^2dx_2dx_1,
\end{eqnarray}
   with the kinematical invariant variables
\begin{eqnarray}
s_1=&(p_Z+p_1)^2&=(p-p_2)^2,\\
s_2=&(p_1+p_2)^2&=(p-p_Z)^2,\\
s_3=&(p_2+p_Z)^2&=(p-p_1)^2,
\end{eqnarray}
  $p_Z$, $p$ and $p_i$ $(i=1,2)$ correspond to the four momenta of the
emitted $Z$ gauge bosons, the Higgs boson and the two leptons in
the final state, and the integration limits are given by:
\begin{eqnarray}
x_{1i}&=&2\sqrt{\xi_Z},\\
x_{1f}&=&1+\xi_Z-4\xi_l,\\
x_{2i}&=&\frac{2-x_1}{2}+\frac{\sqrt{(x_1^2-4\xi_Z)(x_1-1+4\xi_l-\xi_Z)}}{2\sqrt{x_1-\xi_Z-1}},\\
x_{2f}&=&\frac{2-x_1}{2}-\frac{\sqrt{(x_1^2-4\xi_Z)(x_1-1+4\xi_l-\xi_Z)}}{2\sqrt{x_1-\xi_Z-1}},
\end{eqnarray}
with $\xi_Z=m_Z^2/m_h^2$, $\xi_l=m_l^2/m_h^2$,
$s_1=m_h^2(-1+x_1+x_2+\xi_l)$ and $s_2=m_h^2(1-x_1+\xi_Z)$.  The
square of invariant amplitude given in Eq.(\ref{Decay-SM}) will be
expressed in terms of the $hZZ^*$ effective and $Zff$ vertices,
\begin{eqnarray}\label{vertice-1}
g_{hZZ}^{\alpha\beta}&=&\frac{i g
m_Z}{c_W}\left({\cal{G}}_0g^{\alpha\beta}+{\cal G}_{1}\Big[{\cal
F}^gg^{\alpha\beta}+{\cal F}^kk_1^\alpha k_2^\beta\Big]\right),\\
g_{Zff}&=&\frac{ig}{4c_W}\gamma^\mu(f_V+\gamma^5f_A),
\end{eqnarray}
\begin{eqnarray}\label{amplitud}
|{\cal M}|^2&=&\frac{-g^4(f_A^2+f_V^2)}{8 c_W^4m_Z^4}\Big|{\cal G}_{0}+{\cal F}^g {\cal G}_{1}\Big|^2\frac{1}{(s_2-m_Z^2)^2}\Big\{2k_s^4m_f^2(2m_Z^2-s_2)\nonumber\\
&&+m_Z^4 \Big[m_f^4 + m_f^2 (m_h^2 + 3 m_Z^2 - 2 s_1 - s_2)\nonumber \\
&&+ (m_h^2-s_1) (m_Z^2-s_1) + s_2(s_1-2m_Z^2)\Big]\Big\}\nonumber\\
&\equiv&\frac{-g^4(f_A^2+f_V^2)}{8 c_W^4m_Z^4}\Big|{\cal
G}_{0}+{\cal F}^g {\cal G}_{1}\Big|^2 {\cal
K}(s_1,s_2,m_h,m_Z,m_f),
\end{eqnarray}
where we have defined for convenience the function ${\cal
K}(s_1,s_2,m_h,m_Z,m_f)$ and used the
      $hZZ^*$ form factors ${\cal G}_o$, ${\cal G}_1$, ${\cal F}^g$ and ${\cal F}^k$ whose expressions
      are included in Appendix for the SM and the 2HDM. It is important to notice that the form
factor ${\cal F}^k$ does not  contribute to the invariant amplitude given
in Eq.(\ref{amplitud}) because the $Z(p_Z)$ gauge boson is on mass
shell.

\begin{figure}
\begin{center}
\includegraphics[width=2 in]{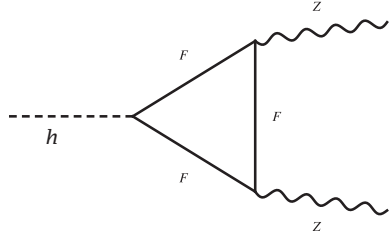}\\
\caption{Generic one-loop fermionic contribution to the $hZZ$ vertex.} \label{h-f-ZZ}
\end{center}
\end{figure}

\begin{figure}
\begin{center}
\includegraphics[width=2 in]{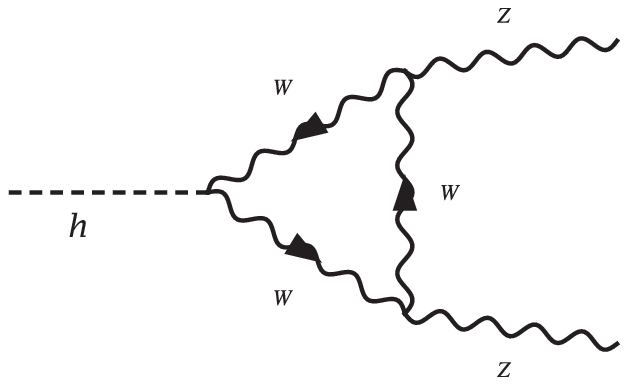}
\includegraphics[width=2 in]{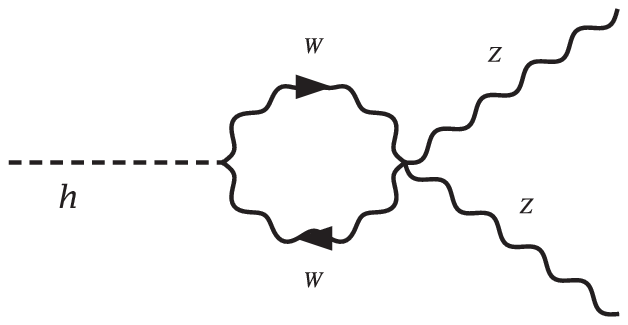}
\caption{Pure $W$\hspace{-.1cm}-$Boson$ contribution to
$hZZ$ vertex.} \label{h-W-ZZ}
\end{center}
\end{figure}

\begin{figure}
\begin{center}
\includegraphics[width=2 in]{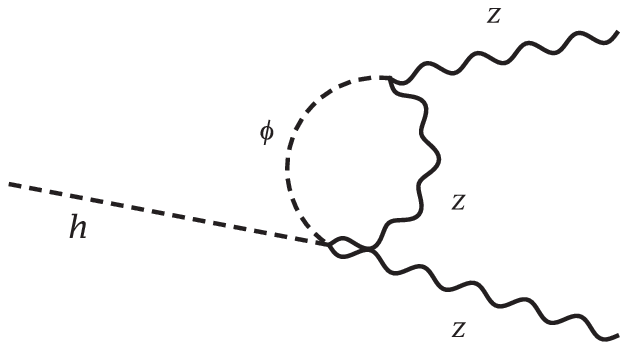}
\includegraphics[width=2 in]{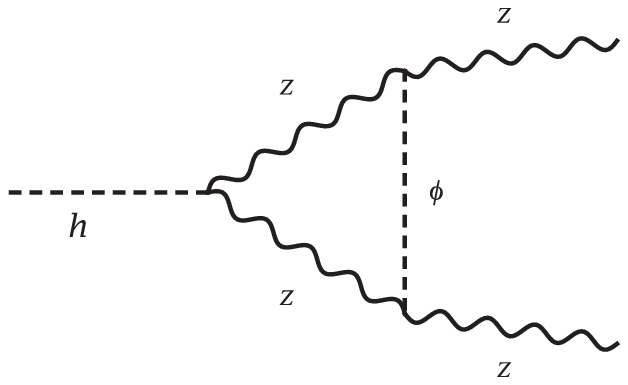}\\
\caption{Scalar contributions to the $hZZ$ vertex that are
independent of the $hhh$ coupling. In the SM  $\phi=h$ and in
2HDM $\phi=\{h,H\}$.} \label{h-ZZ-in}
\end{center}
\end{figure}

\begin{figure}
\begin{center}
\includegraphics[width=2 in]{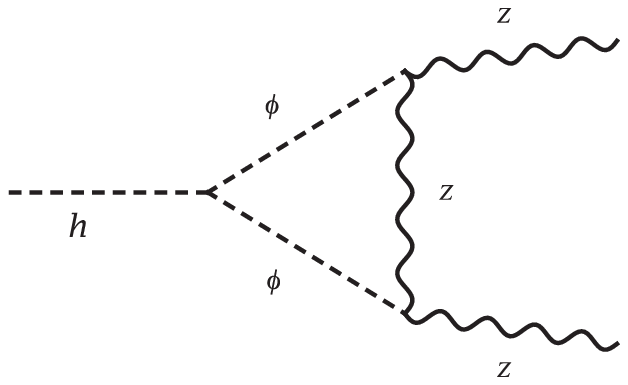}
\includegraphics[width=2 in]{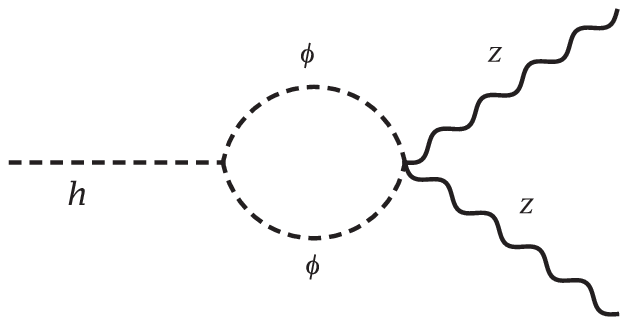}\\
\caption{Scalar contribution to the $hZZ$ vertex that
depends on the  $hhh$ coupling. In the SM  $\phi=h$ and in 2HDM
$\phi=\{h,H\}.$} \label{hhh-ZZ}
\end{center}
\end{figure}

    The contributions to the decay width given in Figures \ref{h-f-ZZ}-\ref{hhh-ZZ} can
    be expressed  in terms of the tree contribution
      (with ${\cal G}_0=1$), the 1-loop term associated to the Feynman diagrams shown in Figures \ref{h-f-ZZ}-\ref{hhh-ZZ},
    and the respective interference term,
\begin{equation}\label{mtree}
|{\cal M}_{SM}|^2_{tree}=\frac{-g^4(f_A^2+f_V^2)}{8 c_W^4m_Z^4}
{\cal K}(s_1,s_2,m_h,m_Z,m_f).
\end{equation}
\begin{eqnarray}
|{\cal M}_{SM}|^2_{1-loop}&=&\frac{-g^4(f_A^2+f_V^2){\cal G}_1^2|{\cal F}_{SM}^g |^2}{8 c_W^4m_Z^4} {\cal K}(s_1,s_2,m_h,m_Z,m_f)\nonumber\\
&=&\frac{-g^4(f_A^2+f_V^2)}{8 c_W^4m_Z^4}\left(\frac{g^2 c_W}{128
\pi^2 k_s^8 m_Z}\right)^2|{{\cal F}^g_f}_{SM}+{{\cal
F}^g_W}_{SM}+{{\cal F}^g_S}_{SM}+{{\cal F}^g_{3h}}_{SM} |^2\nonumber\\
&&\times {\cal
K}(s_1,s_2,m_h,m_Z,m_f),\label{Mloop}
\end{eqnarray}
\begin{equation}
{\cal F}^g_{SM}={{\cal F}^g_f}_{SM}+{{\cal F}^g_W}_{SM}+{{\cal
F}^g_S}_{SM}+{{\cal F}^g_{3h}}_{SM},
\end{equation}
\begin{eqnarray}\label{Mloop2}
{\cal M}_{int}&=&\frac{-g^4(f_A^2+f_V^2)}{4
c_W^4m_Z^4}\left(\frac{g^2 c_W}{128 \pi^2 k_s^8
m_Z}\right)Re[{{\cal F}^g_f}_{SM}+{{\cal F}^g_W}_{SM}+{{\cal
F}^g_S}_{SM}+{{\cal F}^g_{3h}}_{SM}] \nonumber\\
&&\times{\cal
K}(s_1,s_2,m_h,m_Z,m_f).
\end{eqnarray}
The explicit expressions for the SM form factors ${\cal F}^g_{iSM}$
    given in Eqs. (\ref{Mloop}-\ref{Mloop2}) are included in the Appendix.

    The integrated SM decay width takes then the following form
\begin{eqnarray} \label{Decay-SM2}
\Gamma(h_{SM}\to Zll)&=&
\frac{m_{h}}{256\pi^3}\left(\frac{-g^4(f_A^2+f_V^2)}{8c_W^2m_Z^4}\right)\int_{x_{1i}}^{x_{1f}} \int_{x_{2i}}^{x_{2f}}\left[1+\frac{2g^2c_W}{128 \pi^2k_s^8m_Z}\right.\nonumber\\
&&\times Re[{{\cal F}^g_f}_{SM}+{{\cal F}^g_W}_{SM}+{{\cal F}^g_S}_{SM}+{{\cal F}^g_{3h}}_{SM}]\nonumber\\
&&\left.+\left(\frac{g^2c_W}{128 \pi^2k_s^8m_Z}\right)^2 |{{\cal
F}^g_f}_{SM}+{{\cal F}^g_W}_{SM}+{{\cal F}^g_S}_{SM}+{{\cal
F}^g_{3h}}_{SM}|^2\right]\nonumber\\
&&\times {\cal K}(s_1,s_2,m_h,m_Z,m_f)dx_2dx_1.
\end{eqnarray}

\section{2HDM FRAMEWORK}

    The most general Two Higgs Doublet Model (2HDM) potential
    is given by \cite{Gunion}
\begin{eqnarray}
  V ( \Phi_1, \Phi_2 )& = &\ \mu_1^2 ( \Phi_1^{\dag} \Phi_1 ) + \mu_2^2 (
  \Phi_2^{\dag} \Phi_2 ) - \left( \mu_{12}^2 ( \Phi_1^{\dag} \Phi_2 ) + H.c.
  \right) + \lambda_1 ( \Phi_1^{\dag} \Phi_1 )^2  \nonumber\\
  && + \lambda_2 ( \Phi_2^{\dag}
  \Phi_2 )^2 + \lambda_3 ( \Phi_1^{\dag} \Phi_1 ) ( \Phi_2^{\dag} \Phi_2 ) + \lambda_4
  ( \Phi_1^{\dag} \Phi_2 ) ( \Phi_2^{\dag} \Phi_1 )
  \label{poten}\\
  && + \frac{1}{2} \left( \lambda_5 ( \Phi_1^{\dag} \Phi_2 )^2 + \left(
  \lambda_6 ( \Phi_1^{\dag} \Phi_1 ) + \lambda_7 ( \Phi_2^{\dag} \Phi_2 )
  \right) ( \Phi_1^{\dag} \Phi_2 ) + H.c. \right), \nonumber
\end{eqnarray}
    where the two iso-spin Higgs doublets $\Phi_i$, $i=1,2$, receive
    vacuum expectation values $v_1$ and $v_2$, respectively, and
    have hipercharge $+1$. We define  $\tan\beta=v_2/v_1$ and they
    satisfy the relation $v(~246 GeV)=\sqrt{v_1^2+v_2^2}=(\sqrt{2}G_F)^{-1/2}$.
    The parameters $\mu_{12}, \lambda_5, \lambda_6$ and $\lambda_7$  are in general complex numbers, but
    we will set them real in order to maintain CP conservation. A
    discrete symmetry ${\cal Z}_2$ avoids flavor changing neutral currents.
    The terms $\mu_{12}, \lambda_6$ and $\lambda_7$ break explicitly this symmetry but we will
    keep the $\mu_{12}$ term, which violates softly this symmetry, since
    it will play a special role in our calculation of radiative
    corrections and it is required in order to get the decoupling
    limit of the model \cite{Gunion}. Similarly, we will use the full Higgs
    potential Eq. (\ref{poten}) because the $\lambda_6$ and $\lambda_7$ parameters  will
    induce important corrections to the self scalar couplings
    \cite{HernandezSanchez, Cordero-Cid}. This model is called 2HDM-III, it includes two neutral
    scalar fields, $h$ and $H$, and $h$ is the lightest Higgs boson
    associated to the SM. This model predicts also a CP odd scalar
    $A$ and two charged Higgs bosons $H^\pm$.

    \begin{figure}
\begin{center}
\includegraphics[width=2 in]{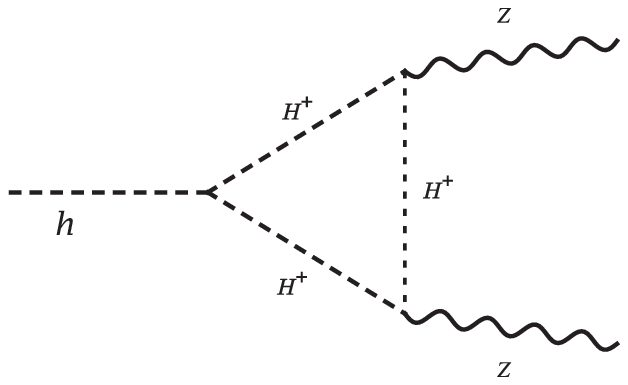}
\includegraphics[width=2 in]{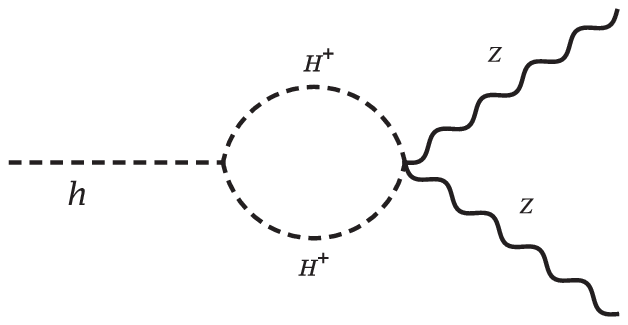}\\
\caption{Higgs charged boson contribution to the $hZZ$
vertex in the 2HDM. } \label{h-Hc-ZZ}
\end{center}
\end{figure}

    The effective vertex $hZZ$ in the 2HDM includes also the Feynman
    diagrams shown in Figure \ref{hhh-ZZ} and two more that are shown in
    Figure \ref{h-Hc-ZZ} for the contributions coming from the virtual exchange
    of charged Higgs bosons. The heavy neutral scalar $H$ also gives
    a contribution in the diagrams of Figures \ref{h-ZZ-in} and \ref{hhh-ZZ}. Even more,
    since the $Zff$ 2HDM couplings are the same as in the SM, the
    invariant decay amplitude is also given by Eq. (\ref{Decay-SM}) with the
    tree-diagram amplitude given by
    \begin{eqnarray}\label{mtree2HDM}
|{\cal M}|^2_{tree}&=&\frac{-g^4(f_A^2+f_V^2)s_{\beta-\alpha}^2}{8 c_W^4m_Z^4} {\cal K}(s_1,s_2,m_h,m_Z,m_f)\nonumber\\
&=&|{\cal M}_{SM}|^{2}_{tree}s_{\beta-\alpha},
\end{eqnarray}
    where $s_{\beta-\alpha}=\sin(\beta-\alpha)$ and $\alpha$ is the angle that defines the
    mixing of the two CP-even scalar bosons. The one-loop amplitude
    has to include the charged Higgs contribution
    \begin{eqnarray}
|{\cal M}|^{2}_{1-loop}&=&\frac{-g^4(f_A^2+f_V^2){\cal G}_1^2|{\cal F}^g |^2}{8 c_W^4m_Z^4} {\cal K}(s_1,s_2,m_\phi,m_Z,m_f)\nonumber\\
&=&\frac{-g^4(f_A^2+f_V^2)}{8 c_W^4m_Z^4}\left(\frac{g^2 c_W}{128 \pi^2 k_s^8 m_Z}\right)^2 {\cal K}(s_1,s_2,m_h,m_Z,m_f)\nonumber\\
&&\times |{\cal F}^{g}_f+{\cal F}^{g}_W+{\cal F}^{g}_S+{\cal
F}^{g}_{3\phi}+{\cal F}^{g}_{H^+} |^2,
\end{eqnarray}
were the function ${\cal{F}}_{H^\pm}^g$ represents the
contribution from the     charged Higgs boson and are included in the Appendix. The functions
${\cal{F}}_{S}^g$
 and ${\cal{F}}_{3\phi}^g$  include     also the contribution coming from
 the heavy neutral Higgs boson $H$ while
the functions ${\cal{F}}_{f}^g$ and ${\cal{F}}_{W}^g$  include a
dependence with     the mixing angles $\beta-\alpha$. The
invariant amplitude for the   interference contribution is now
given by
\begin{eqnarray}
{\cal M}_{int}&=&\frac{-g^4(f_A^2+f_V^2)s_{\beta-\alpha}^2}{4 c_W^4m_Z^4}\left(\frac{g^2 c_W}{128 \pi^2 k_s^8 m_Z}\right){\cal K}(s_1,s_2,m_h,m_Z,m_f)\nonumber\\
&&\times Re[{\cal F}^{g}_f+{\cal F}^{g}_W+{\cal F}^{g}_S+{\cal
F}^{g}_{3\phi}+{\cal F}_{H^+}^{g}],
\end{eqnarray}
and the 2HDM integrated decay width can be expressed by
\begin{eqnarray}
\Gamma(h\to Zll)&=&
\frac{m_{h}}{256\pi^3}\left.\left(\frac{-g^4(f_A^2+f_V^2)}{8c_W^2m_Z^4}\right)\int_{x_{1i}}^{x_{1f}} \int_{x_{2i}}^{x_{2f}}{\cal K}(s_1,s_2,m_h,m_Z,m_f)\right[s_{\beta-\alpha}^2\nonumber\\
&&+\left.\frac{2g^2c_W s_{\beta-\alpha}}{128 \pi^2k_s^8m_Z} Re[{\cal F}^{g}_f+{\cal F}^{g}_W+{\cal F}^{g}_S+{\cal F}^{g}_{3\phi}{\cal F}^{g}_{H^+}]\right.\nonumber\\
&&\left.+\left(\frac{g^2c_W s_{\beta-\alpha}}{128
\pi^2k_s^8m_Z}\right)^2 |{\cal F}^{g}_f+{\cal F}^{g}_W+{\cal
F}^{g}_S+{\cal F}^{g}_{3\phi}{\cal F}^{g}_{H^+}|^2\right]
dx_2dx_1. \label{Anchura-2HDM}
\end{eqnarray}

   \section{ DISCUSION AND CONCLUDING REMARKS}

    In order to determine the sensitivity of the decay width of
    $h \to  ZZ^* \to Zl^+l^-$ to the $hhh$ coupling, we will find convenient
    to define the rate $R$,
    \begin{eqnarray}\label{R}
R&=&\frac{\sigma( gg\to h)\times Br(h\to ZZ^*\to Zl^+l^-)}{\sigma( gg\to h_{SM})\times Br(h_{SM}\to ZZ^*\to Zl^+l^-)_{tree}}\nonumber\\
&\approx&{\cal G}_{htt}^2\frac{ Br(h\to ZZ^*\to
Zl^+l^-)}{Br(h_{SM}\to ZZ^*\to Zl^+l^-)_{tree}},
\end{eqnarray}
    where the factor ${\cal {G}}_{htt} = 1$ for the SM and it varies for each
    type of 2HDM. In the Appendix we include the respective expressions
    of each 2HDM contribution. It is important to stress that
    the main difference in calculating the rate $R$ lies in
    the production cross section for each model that receives
    a dominat contribution from the top-quark one-loop
    Feynman diagram. This contribution is contained in the
    ${\cal G}_{htt}$ factor included in Eq. (\ref{R}).

\begin{figure}[h]
  \centering
  \includegraphics[width=3 in]{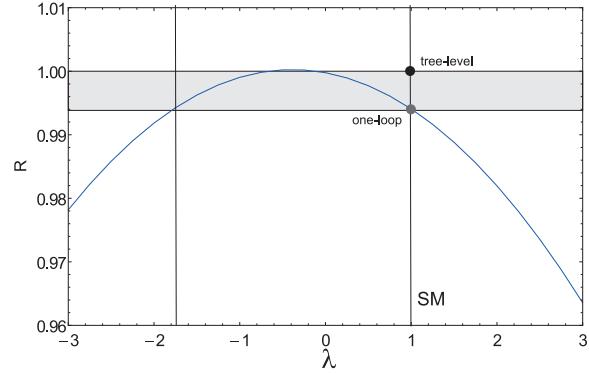}\\
  \caption{Behavior of the ratio $R$ with respect to $\lambda=\lambda_{NP}/\lambda_{SM}$.
   The SM value at tree level is represented by
   a black point and the SM at one loop level  is represented by grey point.}\label{SM-NP}
\end{figure}

We analyze first the behavior of the SM contributions
    associated to the tree- and one-level Feynman diagrams with ${\cal G}_{htt}=1$ in Eq (\ref{R}) and
    \begin{equation}
      R_{SM}=\frac{ Br(h\to ZZ^*\to
Zl^+l^-)}{Br(h_{SM}\to ZZ^*\to Zl^+l^-)_{tree}}.
    \end{equation}
     In Table I are depicted the respective
    contributions coming from each sector of Feynman diagrams.
    The total one-loop correction to the SM rate is below
    the $1\%$ level in agreement with previous results. 
    However, the rate $R_{SM}$ is sensitive to the value used
    for the SM $hhh$ coupling and in Figure \ref{SM-NP} we present the
    dependence of this rate with respect to the normalized
    coupling lambda  in the range $-3< \lambda < +3$. We
    can appreciate that now the radiative correction to
    the SM rate $R_{SM}$ may be as high as $4\%$ with respect to
    the tree-level result. A similar correction will be
    obtained for the 2HDM model.

\begin{center}
\begin{table}
\begin{tabular}{|c|c|}
\hline
 Sector& $R_{SM}=BR_{1-loop}/BR_{tree}$ \\
\hline
total & $0.994$ \\ \hline
Yukawa & $1.0002$ \\ \hline
Gauge & $0.999$ \\ \hline
Scalar (not hhh) & $.999$ \\ \hline
hhh coupling & $0.996$ \\ \hline
\end{tabular}
\caption{The contributions of every sector to rate $R_{SM}$ at one
loop level.} \label{SMtable1}
\end{table}
\end{center}

\begin{figure}[h]
  \centering
  \includegraphics[width=3 in]{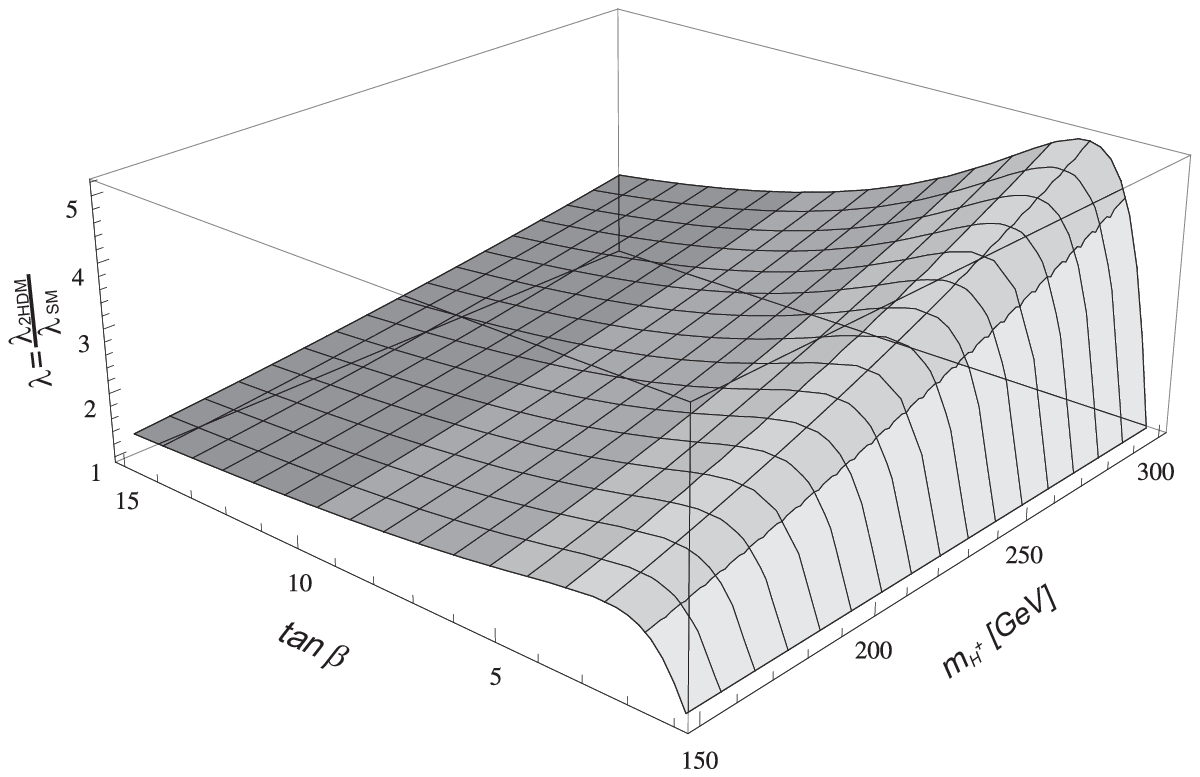}
  \includegraphics[width=3 in]{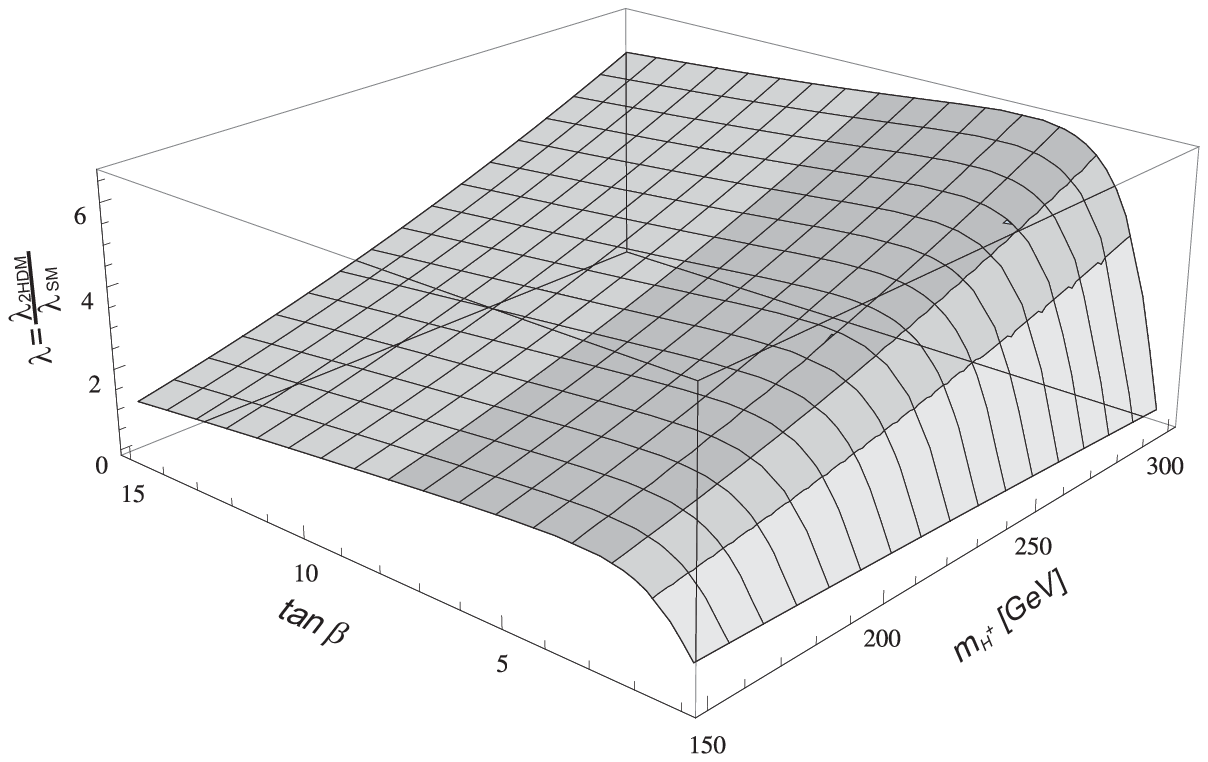}\\
  \includegraphics[width=3 in]{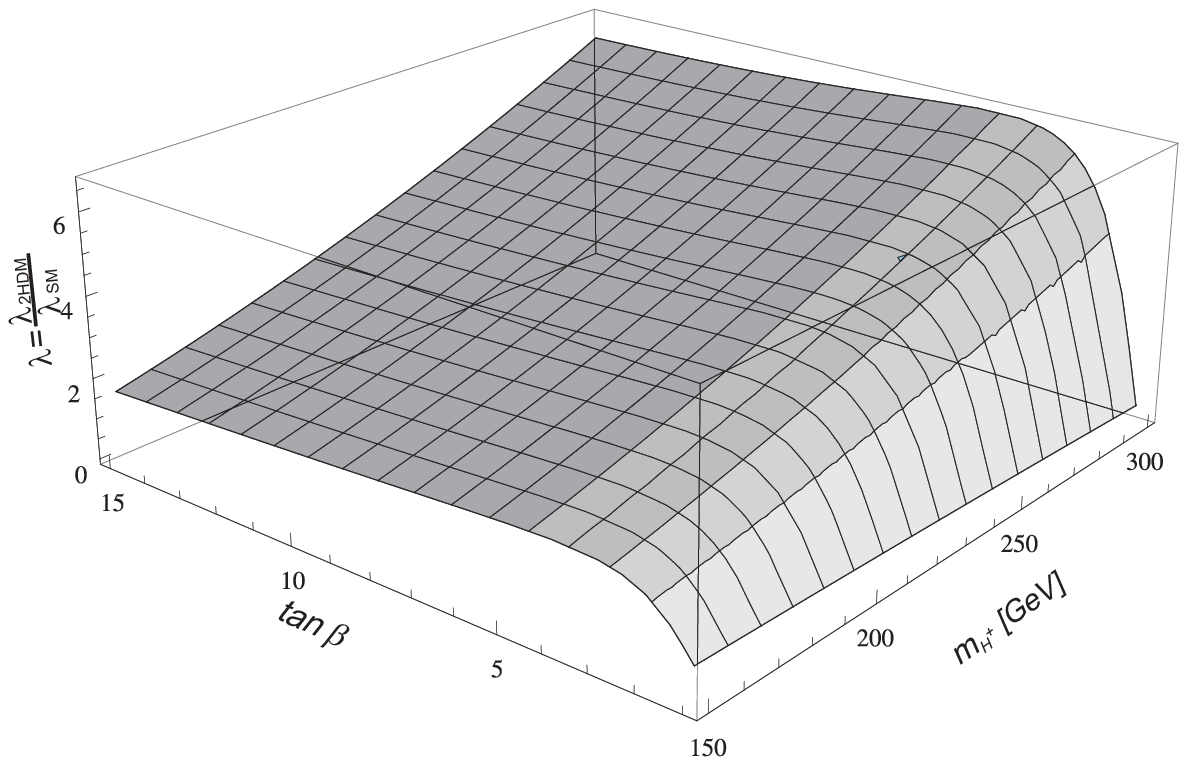}
  \caption{Behavior of lambda ($\lambda=\lambda_{2HDM}/\lambda_{SM}$) around of SM-like scenario ($\beta-\alpha\approx \pi/2$ $\delta=-0.1$):
 a) 2HDM with ${\cal Z}_2$ strict symmetry, b) 2HDM with soft ${\cal Z}_2$ symmetry ($\mu_{12}=200\ GeV$), c) 2HDM without   imposed symmetry ($\mu_{12}=200\ GeV,\ \lambda_6=\lambda_7=1$). }\label{2hdm-lambda}
\end{figure}

 The 2HDM presents strong modifications for the Higgs self
    coupling: it is rather sensitive to the mixing angles
    and other couplings. This is the reason for presenting
    our results for the radiative corrections in three versions
    of the 2HDM: with ${\cal Z}_2$ exact symmetry, with only soft
    violations of the ${\cal Z}_2$ symmetry, and for the most general
    2HDM. We present in Figure \ref{2hdm-lambda} the dependence of the
    normaized coupling $\lambda$ with respect to $\tan \beta$ and
    $m_{H^+}$. There is a strong dependence of $\lambda$ for each
    of the three 2HDM parameters, with typical increases of
    order $\lambda \sim 4$.

\begin{figure}[h]
  \centering
  \includegraphics[width=3 in]{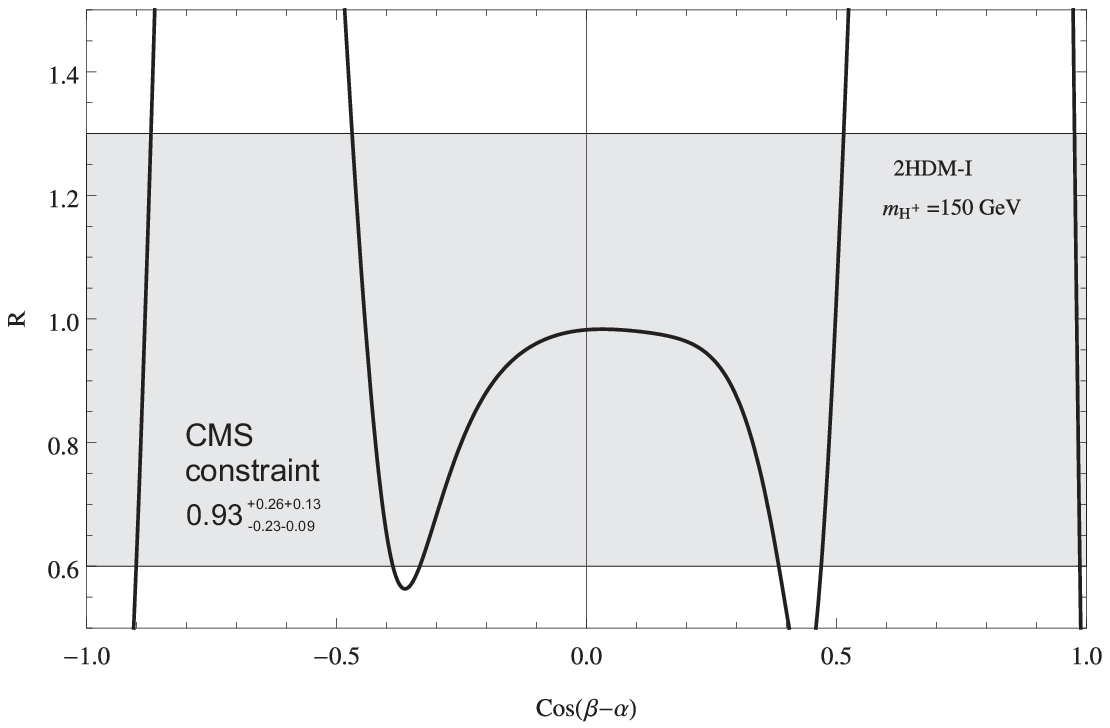}
\includegraphics[width=2.1 in]{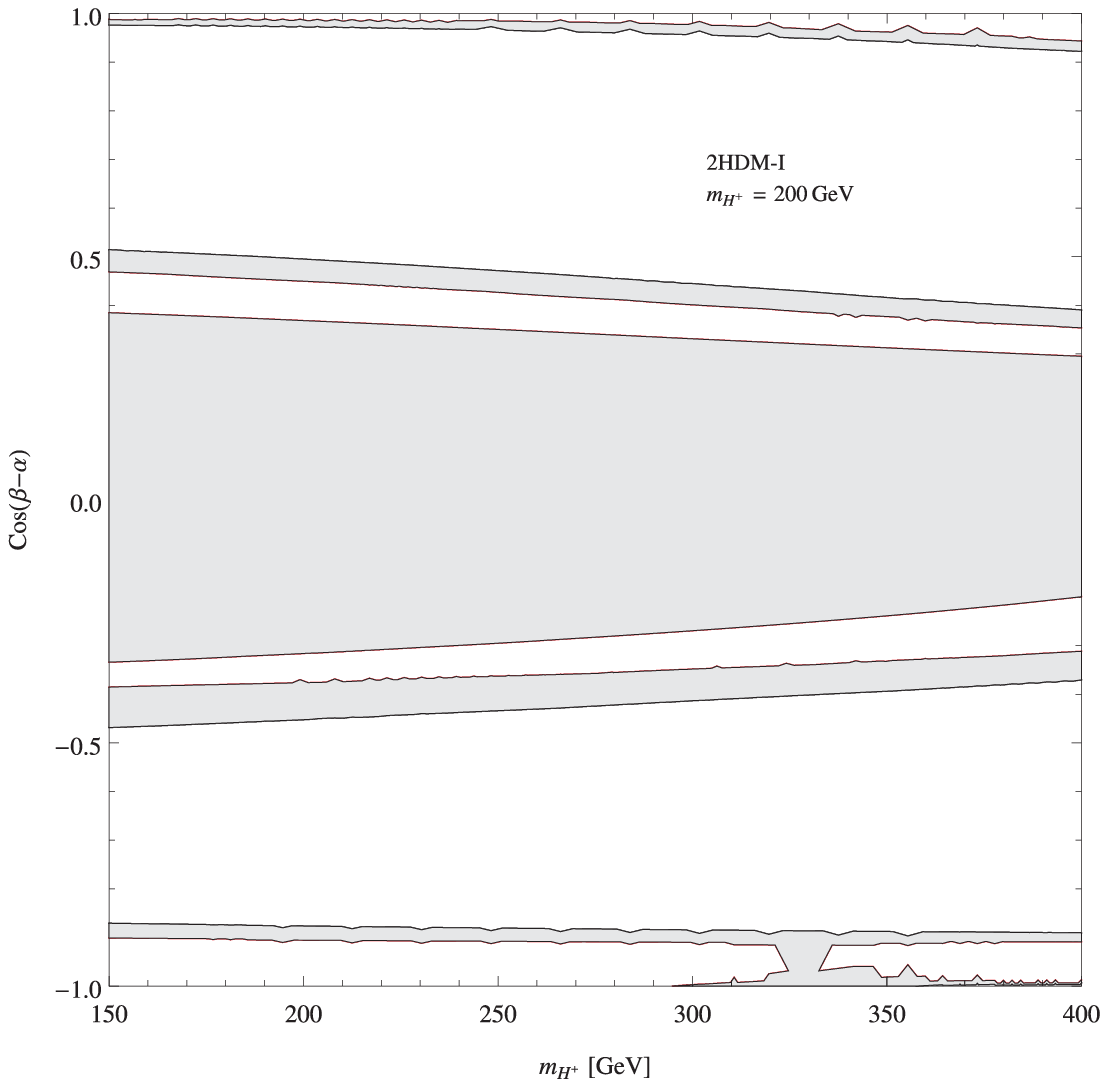}\\
  \includegraphics[width=3 in]{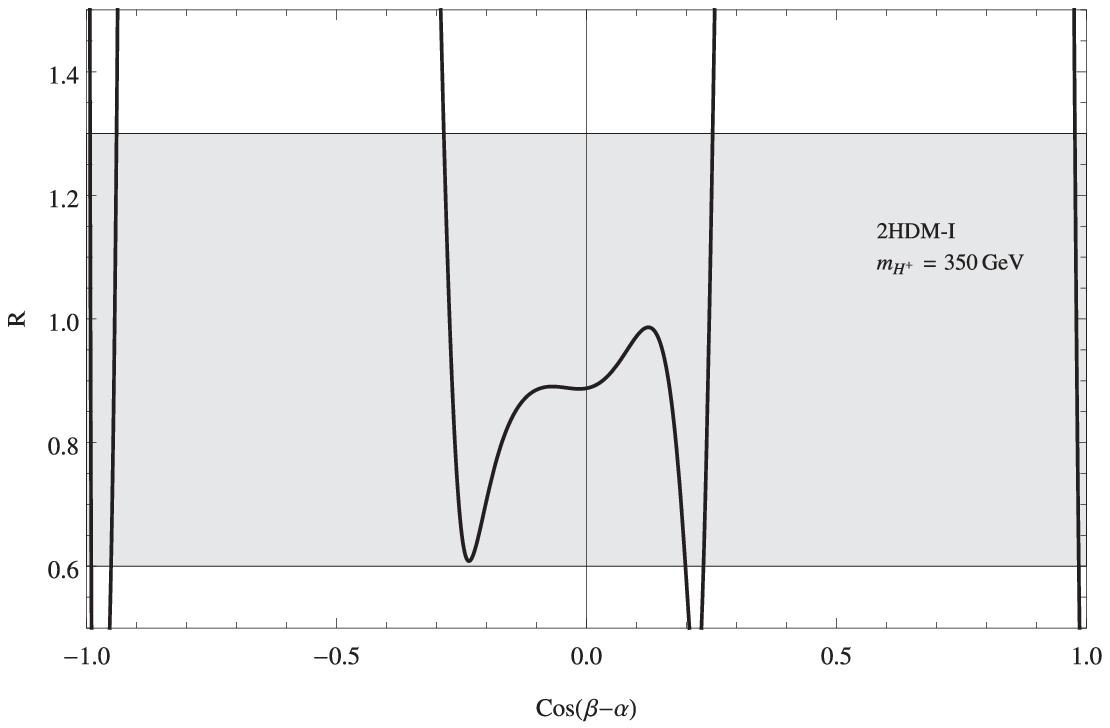}
\includegraphics[width=2.1 in]{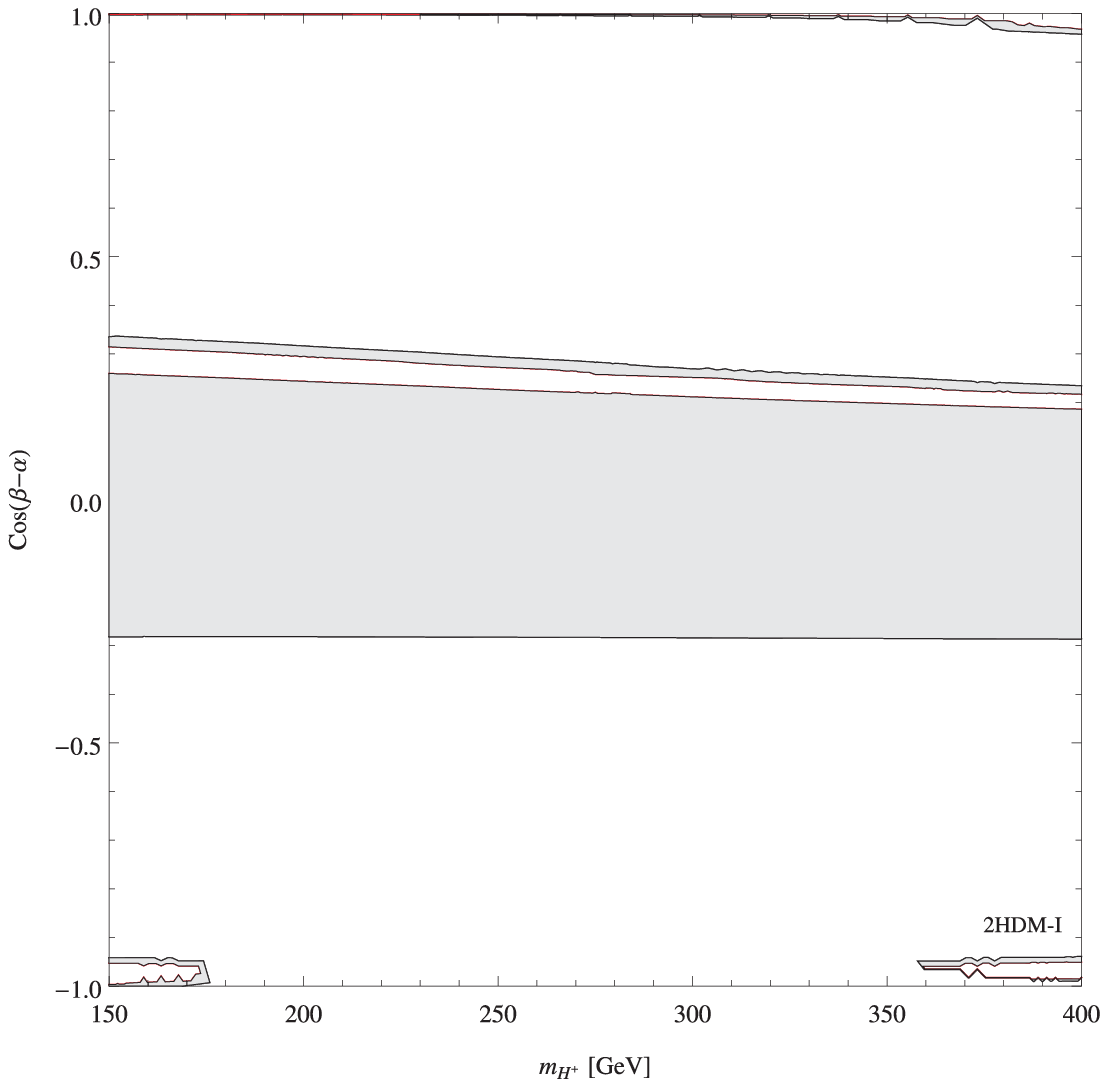}\\
\includegraphics[width=3 in]{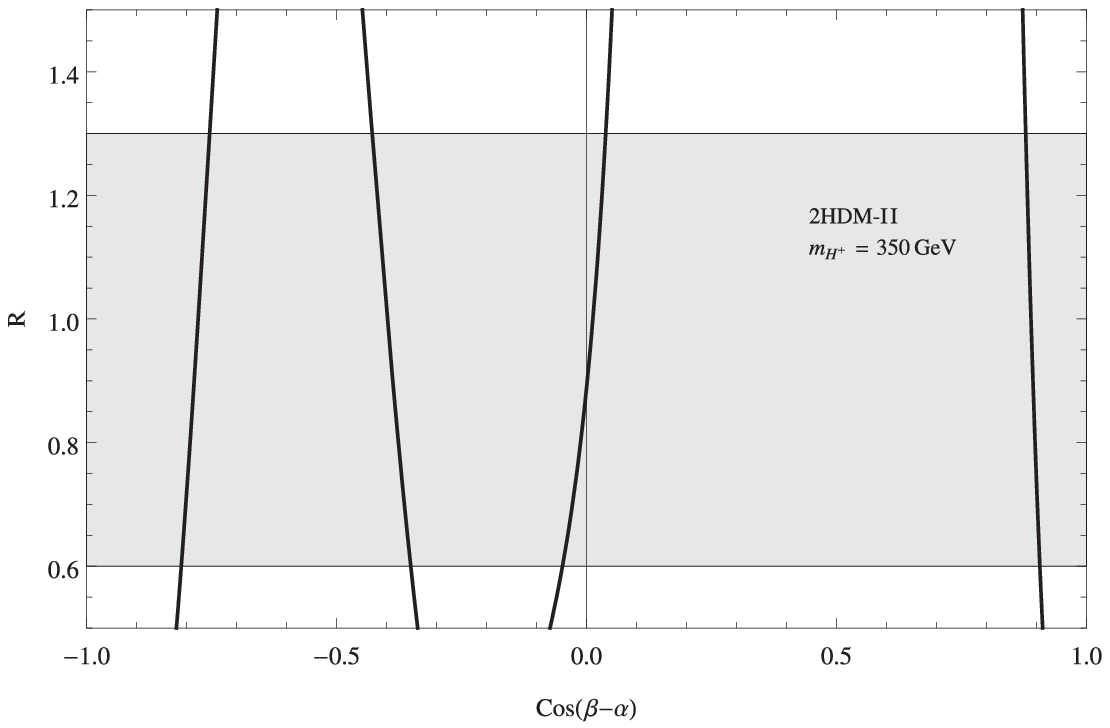}
\includegraphics[width=2.1 in]{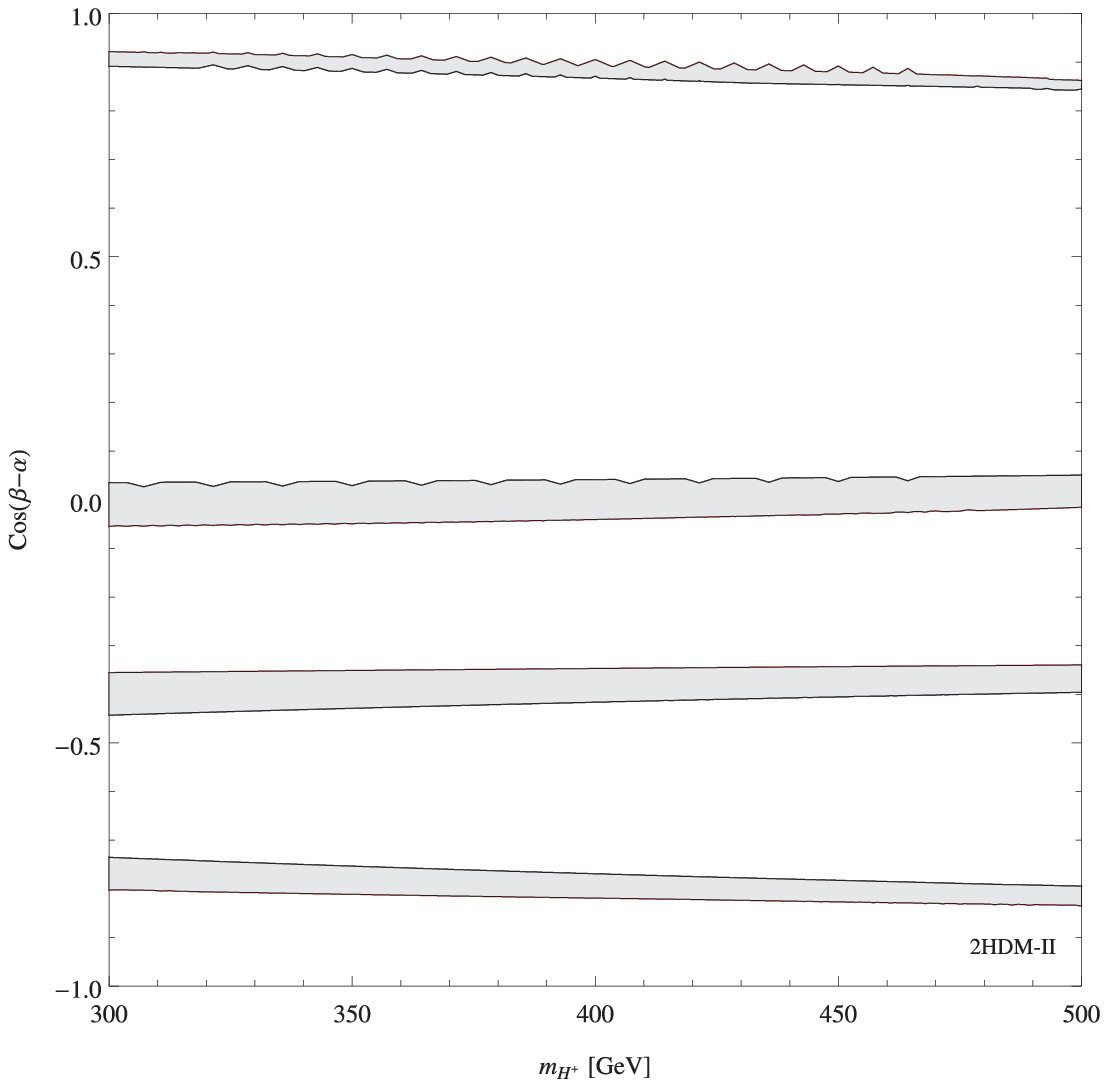}
\\
\caption{The grey areas correspond to the allowed regions
            imposed to the rate R by the LHC on the 2HDM mixing
            angles and the charged Higgs boson mass \cite{Chatrchyan}. In the
            left plots, we depict the variation of $R$ with
            respect to the two types of 2HDM with ${\cal Z}_2$ symmetry
            and the right plots we include the respective constrains
            used in the $m_{H^+}-\cos(\beta-\alpha)$ space.}
\label{2hdm-Z2}
\end{figure}

\begin{figure}[h]
  \centering
  \includegraphics[width=3 in]{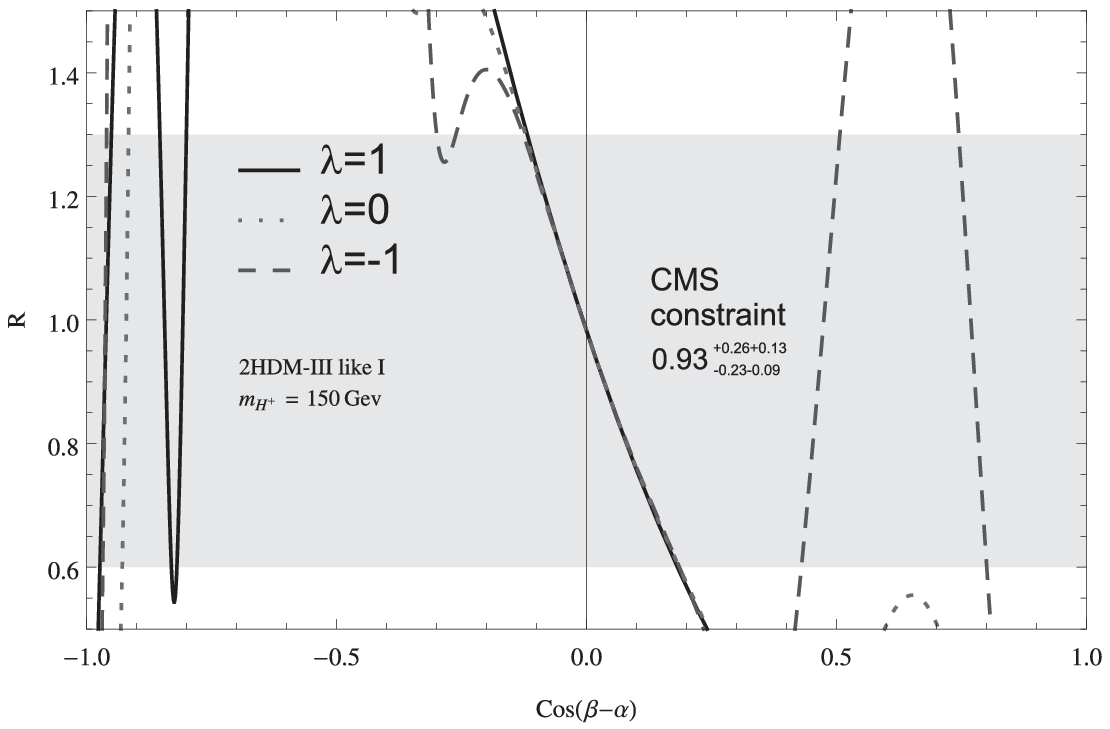}
\includegraphics[width=2.1 in]{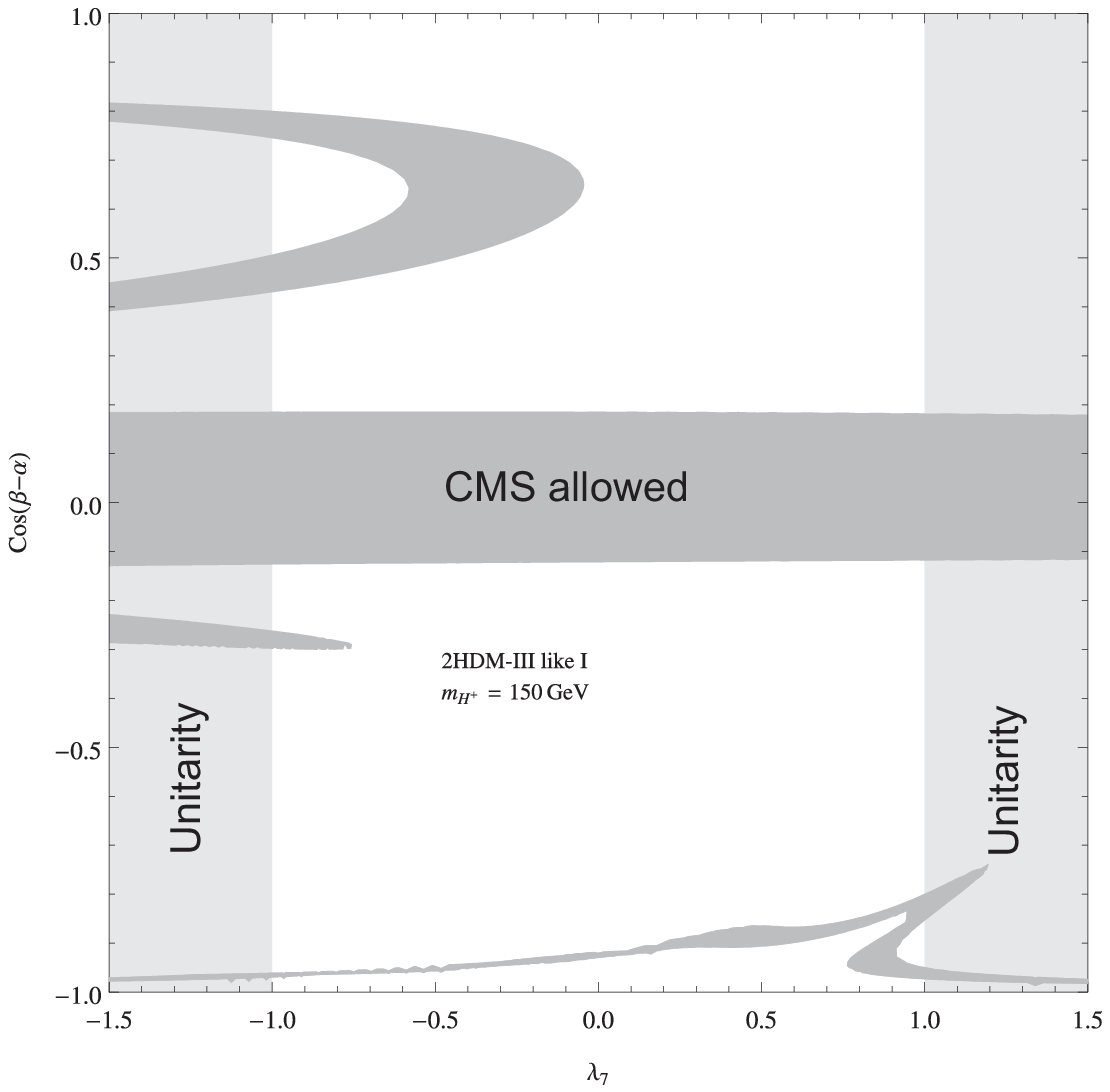}\\
  \includegraphics[width=3 in]{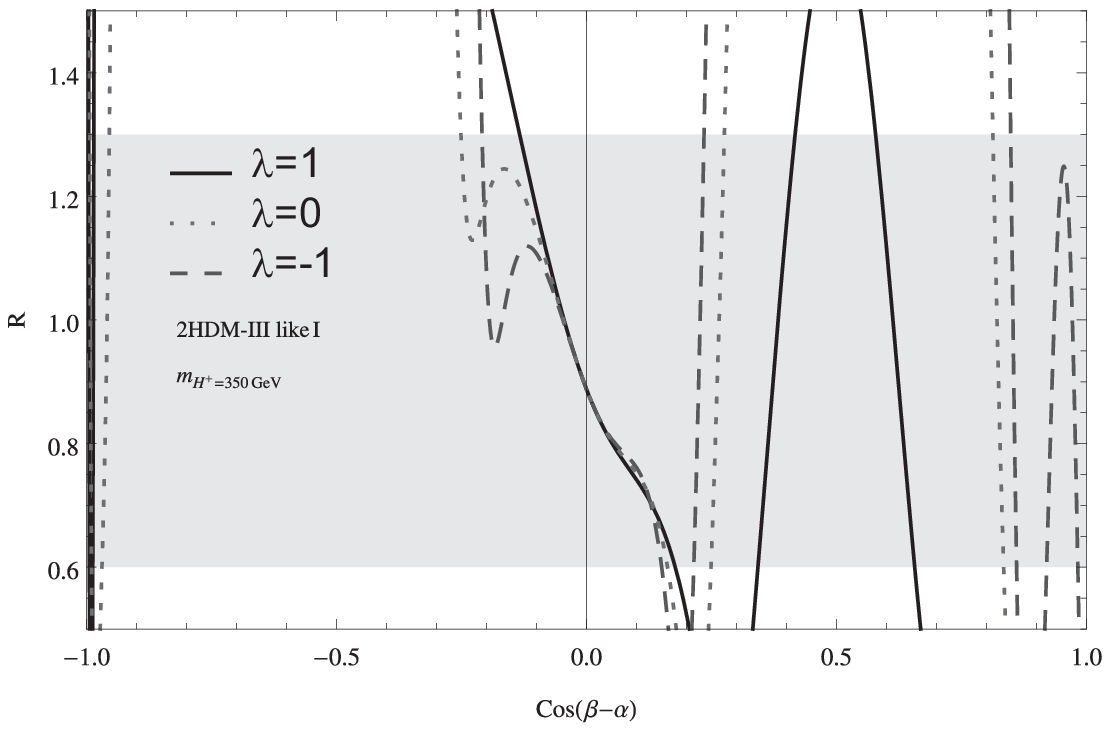}
\includegraphics[width=2.1 in]{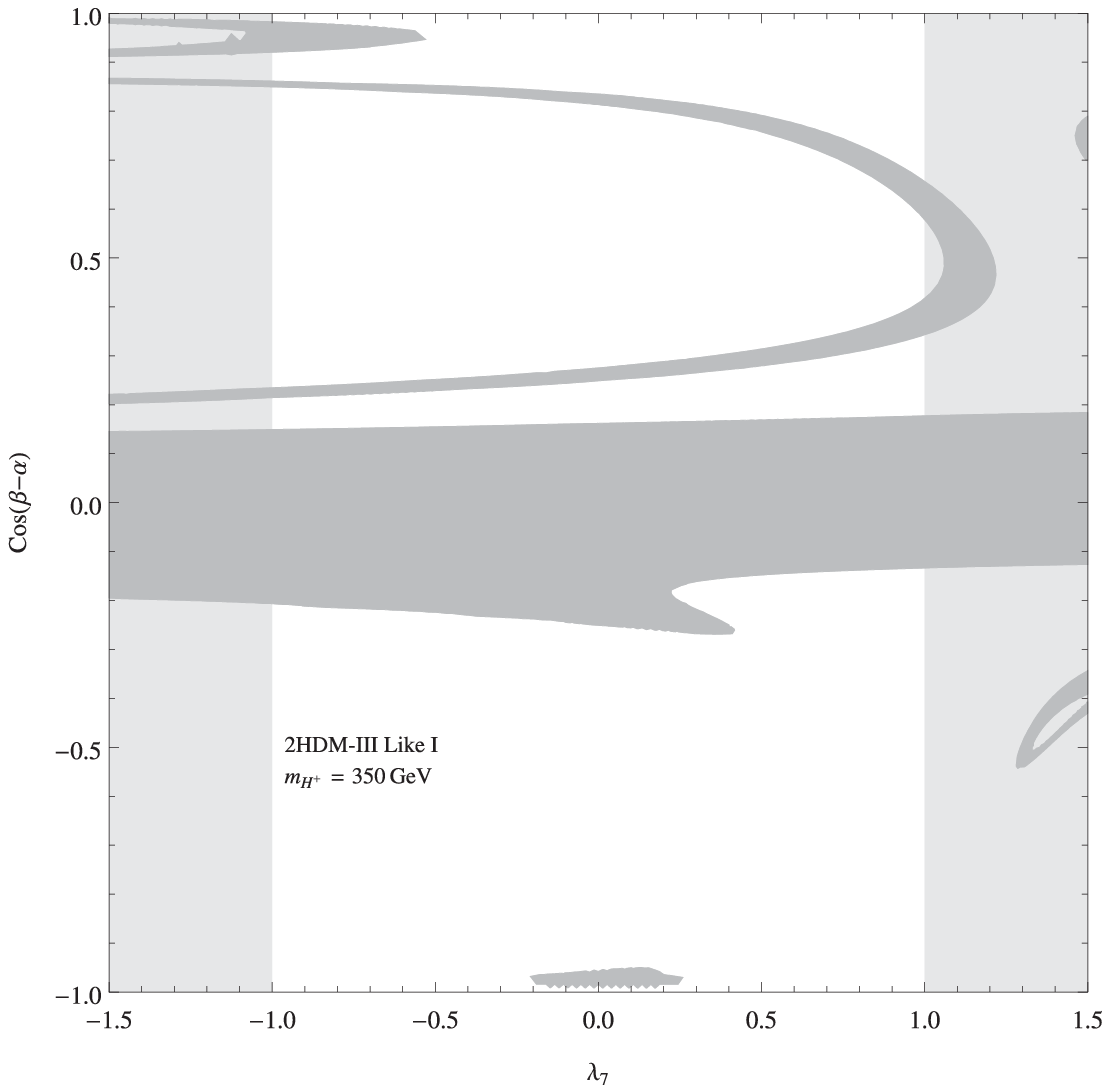}\\
\includegraphics[width=3 in]{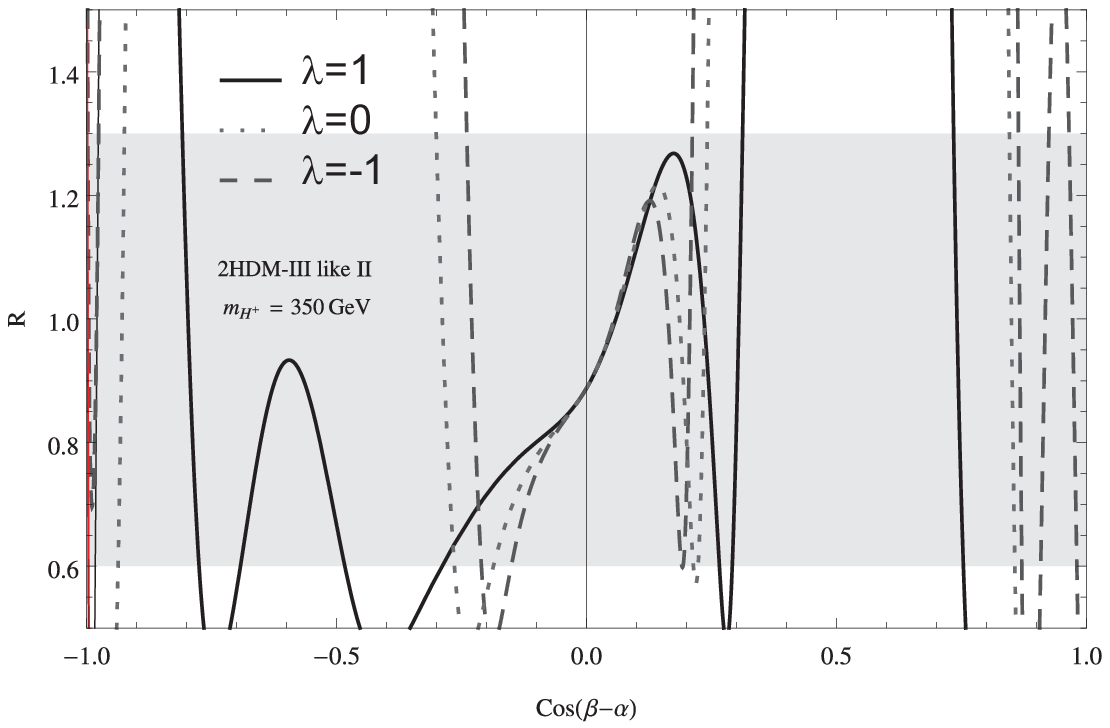}
\includegraphics[width=2.1 in]{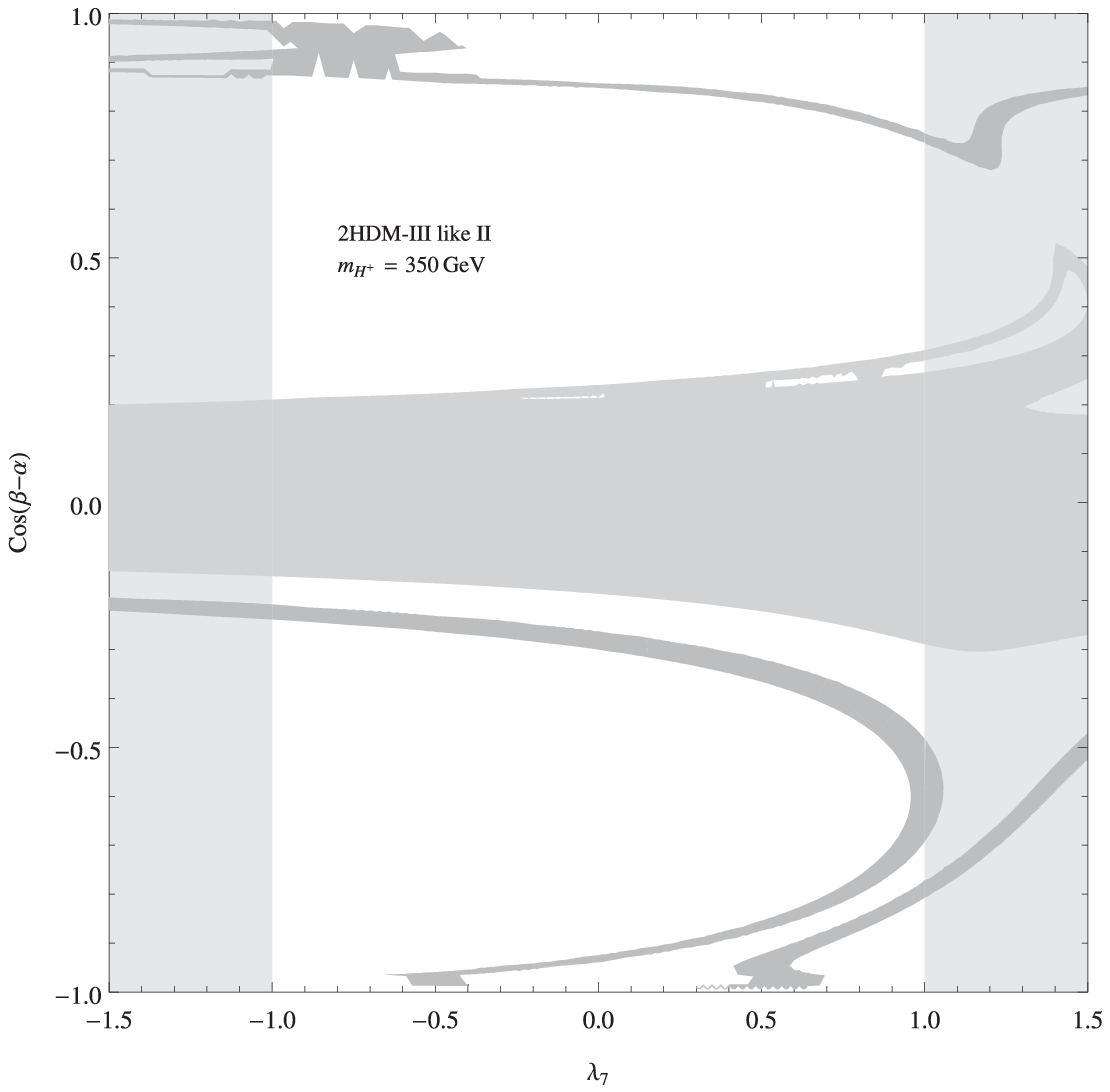}
\\
\caption{The grey areas correspond to the allowed regions
            imposed to the rate R by the LHC on the 2HDM mixing
            angles and the charged Higgs boson mass \cite{Chatrchyan}. In the
            left plots, we depict the variation of $R$ with
            respect to the two types of parametrizations forthe 2HDM-III
            and the right plots we include the respective constrains
            used in the $m_{H^+}-\cos(\beta-\alpha)$ space.}
\label{2hdm-III}
\end{figure}

In Table \ref{2hdmtable1} we depict the respective contributions coming
    from the radiative corrections induced on $R_{2HDM}$ by
    the different sectors of the 2HDM with ${\cal Z}_2$ symmetry.
    The enhancement of the Rate $ R_{2HDM}$ can as high as $4\%$.
    The main difference with respect to the SM result comes
    from the charged Higgs loops and the neutral Higgs bosons
    $H$ and $A$. Accordingly, the enhancement could be even larger
    in the other two versions of the 2HDM, of order $30\%$ due to the weak limits imposed by the LHC data on the
    mixing angles and masses of the extra Higgs bosons as
    it  is shown in Figures \ref{2hdm-Z2} and \ref{2hdm-III}. This flexibility
    induces a large radiative correction to the rate $R_{2HDM}$
    via the corrected triple Higgs bosons couplings of these
    models.

\begin{center}
\begin{table}
\begin{tabular}{|c|c|c|c|}
\hline
 Sector& $R=BR_{1-loop}^{2HDM-I}/BR_{tree}^{SM}$  & $R=BR_{1-loop}^{2HDM-I}/BR_{tree}^{SM}$&$R=BR_{1-loop}^{2HDM-II}/BR_{tree}^{SM}$ \\
\hline
total & $0.961$ & $0.885$ & $0.682$\\ \hline
Yukawa & $0.985$ & $0.987$ & $0.775$\\ \hline
Gauge & $0.984$ & $0.987$ & $0.774$\\ \hline
Scalar (not $\phi\phi\phi$) & $0.984$ & $0.986$  & $0.775$\\ \hline
Higgs charged& $0.985$  & $0.987$  & $0.775$\\ \hline
triple Higgs & $0.966$ & $0.896$ & $0.692$\\ \hline
\end{tabular}
\caption{The contributions of every sector to the rate $R_{2HDM}$ with ${\cal Z}_2$ symmetry. The firts column represents 2HDM-I with $m_{H^+}=150\ GeV$, $\tan\beta=5$ and $\alpha=\beta-\pi/2-0.1$, the second column is for 2HDM-I with $m_{H^+}=350\ GeV$, $\tan\beta=5$ and $\alpha=\beta-\pi/2-0.1$ and the third column represent the 2HDM-II with $m_{H^+}=150\ GeV$, $\tan\beta=5$ and $\alpha=\beta-\pi/2-0.01.$ } \label{2hdmtable1}
\end{table}
\end{center}

 In conclusion, we have presented the sensitivity expected
    in the decay rate of the Higgs decay $h\to ZZ^*\to Zl^+l^-$
    to the radiative corrections induced by the triple
    Higgs boson coupling shown in Figures \ref{hhh-ZZ} and \ref{h-Hc-ZZ}. While
    in the SM this effect is below the $1\%$ level, there
    could be an enhancement of one order of magnitude if the
    $hhh$ coupling differs from the SM expectation of
    the Higgs self coupling. On the other hand, in the
    three versions studied of the 2HDM, we have found that
    the radiative corrections to the Higgs decay rate could
    be much larger due to the weak limits imposed to the
    mixing angles and masses of the new Higgs bosons.
    In the latter case, a new window could be opened to
    test the predictions of the 2HDM if the Higgs decay
    width could be measured with an accuracy below the
    $5\%$ level.

    \section*{AKCNOWLEDGMENTS}

    We appreciate the support from CONACyT (Mexico) and
    useful discussions with F. Larios.

\appendix

\section*{Appendix}

   We have used in our calculations the Feynman rules
    corresponding to the 2HDM-III which are given in detail
    in Refs. \cite{HernandezSanchez,Cordero-Cid}. The gauge boson vector couplings were
    used in the unitary gauge and the respective Yukawa
    couplings are determined by the four zeros texture version
    of 2HDM \cite{HernandezSanchez:2012eg}. The respective 
     factors
     ${\cal G}_{\phi ff}$, ${\cal G}_{h WW}$ and    ${\cal G}_{h\phi_i \phi_j}$
     required for the one-loop calculations are depicted in
    Tables \ref{yukawa1} and \ref{couplings1}.

\begin{table}
\centering
\begin{tabular}{|c|c|c|c|}
 \hline
 ${\cal G}_{h ff}$&$ll$&$dd$&$uu$\\
 \hline
2HDM-III Like I&$\frac{c_\alpha}{s_\beta}+\frac{\chi_{ll}c_{\beta-\alpha}}{\sqrt{2}s_\beta}$&
 $\frac{c_\alpha}{s_\beta}+\frac{\chi_{dd}c_{\beta-\alpha}}{\sqrt{2}s_\beta}$&
 $\frac{c_\alpha}{s_\beta}-\frac{\chi_{uu}c_{\beta-\alpha}}{\sqrt{2}s_\beta}$\\
\hline
 2HDM-III Like
II&$\frac{-s_\alpha}{c_\beta}+\frac{\chi_{ll}c_{\beta-\alpha}}{\sqrt{2}c_\beta}$&
$\frac{-s_\alpha}{c_\beta}+\frac{\chi_{dd}c_{\beta-\alpha}}{\sqrt{2}c_\beta}$&
$\frac{c_\alpha}{s_\beta}-\frac{\chi_{uu}c_{\beta-\alpha}}{\sqrt{2}s_\beta}$\\
\hline
\end{tabular}
\caption{Yukawa couplings for the neutral Higgs boson $h$ in the
2HDM-III. The particular cases for $SM$ and $2HDM$ with ${\cal
Z}_2$ symmetry are achieved when $\beta-\alpha=\pi/2$, or
$\chi_{ff}=0$, respectively.} \label{yukawa1}
\end{table}

\begin{table}
\begin{center}
\begin{tabular}{|c|l|}
\hline
 Factor coupling&Factor function\\
\hline ${\cal G}_{hH^+H^-}$&$\frac{-1}{16 g^2 m_W^2}\Big\{16
g^2\mu_{12}^2\frac{c_{\alpha+\beta}}
{s_{2\beta}^2}-2g^2m_h^2\frac{c_{\alpha-3\beta}+3c_{\alpha+\beta}}{s_{2\beta}}-8m_W^2\lambda_6
\frac{c_{\alpha-\beta}}{s_{\beta}^2}+8m_W^2\lambda_7\frac{c_{\alpha-\beta}}{c_\beta^2}$\\
&$+\frac{g^2m_{H^+}^2}{c_\beta}\Big(s_{\alpha-2 \beta}+3 s_{2
\alpha-\beta} -s_{\alpha+2\beta}+s_{2 \alpha+3
\beta}-s_{\alpha+4\beta}+s_{\alpha}-3 s_{\beta}+s_{3
\beta}\Big)\Big\}$\\
 \hline
 ${\cal G}_{hhh}$&$2 m_h^2\frac{(c_{3 \alpha - \beta} + 3 c_{\alpha +\beta} )}{s_{2\beta}}
  +2 m_{H^+}^2 s_{2 \alpha}\frac{(-c_{2 \alpha + \beta} + c_{\alpha + 2\beta}
   + c_\alpha - c_\beta)}{c_\beta} -16 \mu_{12}^2  \frac{c_{\alpha+\beta}
 c_{\alpha-\beta}^2}{s_{2\beta}^2}$\\
 &$+8 m_W^2 \lambda_6 \frac{c^3_{\alpha-\beta}}{g^2s_{\beta}^2}-8 m_W^2 \lambda_7
 \frac{c^3_{\alpha-\beta}}{g^2c_{\beta}^2}$\\
 \hline
 ${\cal G}_{hhH}$&$\frac{1}{3}\Big\{4 m_H^2 s_{2\alpha} \frac{ c_{\alpha-\beta}}{s_{2\beta}} +8 m_h^2 s_{2\alpha}
  \frac{c_{\alpha-\beta}}{s_{2\beta}}+2m_{H^+}^2
  \frac{(c_\alpha-c_\beta)(c_{\beta-\alpha}+3c_{3\alpha+\beta})}{c_\beta}$\\
  &$-4 \mu_{12}^2 \frac{(s_{\alpha - 3 \beta} + 3 s_{3 \alpha - \beta} + 2 s_{\alpha +\beta})}{s_{2\beta}^2}+24 m_W^2
  \lambda_6  \frac{s_{\alpha - \beta} c^2_{\alpha - \beta}}{g^2s_\beta^2}-24 m_W^2 \lambda_7  \frac{s_{\alpha - \beta}
  c^2_{\alpha - \beta}}{g^2c_\beta^2}$\\
  \hline
  ${\cal G}_{hHH}$&$\frac{1}{3}\Big\{4 m_h^2 s_{2\alpha} \frac{ s_{\alpha-\beta}}{s_{2\beta}} +8 m_H^2 s_{2\alpha}
  \frac{s_{\alpha-\beta}}{s_{2\beta}}+2m_{H^+}^2
  \frac{(c_\alpha-c_\beta)(s_{\beta-\alpha}+3s_{3\alpha+\beta})}{c_\beta}$\\
  &$-4 \mu_{12}^2 \frac{(c_{\alpha - 3 \beta} - 3 c_{3 \alpha - \beta} + 2 c_{\alpha +\beta})}{s_{2\beta}^2}+24 m_W^2 \lambda_6
   \frac{s^2_{\alpha - \beta} c_{\alpha - \beta}}{g^2s_\beta^2}-24 m_W^2 \lambda_7  \frac{s^2_{\alpha - \beta} c_{\alpha -
  \beta}}{g^2c_\beta^2}$\\
  \hline
  ${\cal G}_{hWW}$&$s_{\beta-\alpha}$\\
  \hline
  ${\cal G}_{HWW}$&$c_{\beta-\alpha}$\\
  \hline
\end{tabular}
\end{center}
\caption{ Scalar and vector boson couplings in the 2HDM-III.  }
\label{couplings1}
\end{table}

   We include bellow also the explicit expressions for the one-loop
    contributions obtained for the fermion loops (Figure \ref{h-f-ZZ}),
    the $W$ boson loops (Figure \ref{h-W-ZZ}), the scalar-$Z$ boson loops
    (Figure \ref{h-ZZ-in}) and the charged and neutral Higgs boson loops
    (Figures \ref{hhh-ZZ} and  \ref{h-Hc-ZZ}).

Contribution for the fermion loop (Figure \ref{h-f-ZZ}):
\begin{eqnarray}\label{ffg}
{\cal F}_{f}^g&=&\sum_f\frac{ -N_c^f m_f^2{\cal
G}_{hff}}{c_W^2m_W}
\left\{k_s^8(f_A^2+f_V^2)+k_s^4\Big(k_1^2\Big[(f_A^2+f_V^2)[k_1\cdot k_2]^2+k_s^4(f_A^2-f_V^2)\Big]\right.\nonumber\\
&&-2k_s^4\Big[m_f^2(f_A^2-f_V^2)+f_V^2k_1\cdot k_2\Big]+(f_A^2+f_V^2)[k_1\cdot k_2]^3\Big)C_0(k_1,k_2)\nonumber\\
&&+k_s^4\Big((f_A^2+f_V^2)k_1\cdot k_2[k_1^2+k_1\cdot k_2]+k_s^4f_V^2\Big)B_0(p,k_1)\nonumber\\
&&\left.-k_s^8f_A^2\Big(B_\mu(k_1,\mu)+B_\mu(p,\mu)\Big)\right\}+\{k_1\leftrightarrow
k_2\}, \label{ampli-zz}
\end{eqnarray}
the $f_A$ and $f_V$ represent the axial and vectorial
contribution, respectively. On the other hand, we have introduced a
short notation for the Passarino-Veltman functions
\begin{eqnarray}
C_0(k_i,k_j)&=&C_0(k_i^2,k_j^2,(k_i+k_j)^2,m^2,m^2,m^2),\label{c0}\\
B_0(k_i,k_j)&=&B_0(k_i^2,m^2,m^2)-B_0(k_j^2,m^2,m^2),\label{b0}\\
B_\mu(k_i,\mu_R)&=&B_0(k_i,m^2+\mu_R^2,m^2+\mu_R^2)-B_0(0,\mu_R^2,\mu_R^2),\label{bmu}
\end{eqnarray}
where $B_0(k_i,k_j)$ is a finite contribution, and
$B_\mu(k_i,\mu_R)$ is a renormalized contribution  \cite{Pittau:2012zd}.

Contribution for the $W$-boson loop (Figure \ref{h-W-ZZ}):
\begin{eqnarray}\label{fwg}
{\cal F}_W^g&=&\frac{c_W^2{\cal G}_{hWW}}{9m_W^5}\Big(2 k_s^8
\Big\{-18 k_1^4 m_W^2+k_1^2(-36m_W^2 k_1\cdot k_2+11[k_1\cdot
k_2]^2-11k_s^4)\nonumber\\&&-9(k_s^4-4m_W^2)k_1\cdot k_2 -31
m_W^2[k_1\cdot k_2]^2+9[k_1\cdot
k_2]^3+31k_s^4m_W^2+108m_W^6\Big\}\nonumber\\
&&+9k_s^4\Big\{2k_1^6m_W^2(k_s^4-[k_1\cdot k_2]^2)
+k_1^4\Big[-2k_s^4(k_1\cdot k_2-2m_W^2)^2\nonumber\\&&+[k_1\cdot
k_2]^3(k_1\cdot
k_2-8m_W^2)+k_s^8\Big]+4k_1^2\Big[-2(k_s^4-m_W^4)[k_1\cdot
k_2]^3\nonumber\\
&&+3m_W^2(k_s^4+2m_W^4)[k_1\cdot
k_2]^2+k_s^4(k_s^4+2m_W^4)k_1\cdot k_2-3m_W^2[k_1\cdot
k_2]^4\nonumber\\
&&+[k_1\cdot k_2]^5-6k_s^4m_W^6\Big]-k_s^8(2m_W^2k_1\cdot k_2-5[k_1\cdot k_2]^2+12m_W^4)\nonumber\\
&&+k_s^4(-72m_W^6k_1\cdot k_2+4m_W^4[k_1\cdot k_2]^2+8m_W^2[k_1\cdot k_2]^3-7[k_1\cdot k_2]^4+48m_W^8)\nonumber\\
&&+[k_1\cdot k_2]^3(8m_W^4k_1\cdot k_2-6m_W^2[k_1\cdot k_2]^2+3[k_1\cdot k_2]^3+24m_W^6)-k_s^{12}\Big\}C_0(k_1,k_2)\nonumber\\
&&-12 k_s^8 m_W^2 (k_s^4-4 k_2^2 m_W^2)B_0(k_1,0)
-9 k_s^4 \Big\{k_1^4 (-k_s^4 k_1\cdot k_2-6 m_W^2 [k_1\cdot k_2]^2\nonumber\\
&&+[k_1\cdot k_2]^3+2 k_s^4 m_W^2)+[k_1\cdot k_2]^3 (-6 k_2^2 m_W^2-4 k_s^4+8 m_W^4)\nonumber\\
&&+[k_1\cdot k_2]^2 \big[4 m_W^2 (k_s^4+6 m_W^4)-2 k_2^2 (k_2^2 m_W^2+k_s^4)\big]\nonumber\\
&&+k_1^2 k_1\cdot k_2 \big[k_1\cdot k_2 (3 k_1\cdot k_2 (k_1\cdot k_2-2 m_W^2)-2 k_s^4+8 m_W^4)+6 m_W^2 (k_s^4+4 m_W^4)\big]\nonumber\\
&&+k_s^4 k_1\cdot k_2 (6 k_2^2 m_W^2+k_s^4-8 m_W^4)+(k_2^2-6 m_W^2) [k_1\cdot k_2]^4-2 k_1^6 m_W^2 k_1\cdot k_2\nonumber\\
&&+3 [k_1\cdot k_2]^5+2 k_s^4 m_W^2 (k_2^4-12 m_W^4)\Big\}B_0(k_1,p)+12 k_s^8 m_W^2 [k_1\cdot k_2]^2 B_0(p,0)\nonumber\\
&&-3 k_s^8 \Big\{6 k_s^4 k_1\cdot k_2+k_1^2 (5 k_s^4-8 [k_1\cdot k_2]^2)-6 [k_1\cdot k_2]^3+k_s^4 (3 k_2^2+2 m_W^2)\Big\}B_\mu(k_1,\mu)\nonumber\\
&&+6  k_s^8 m_W^2 (7 [k_1\cdot k_2]^2-6
k_s^4)B_\mu(p,\mu)\Big)+\{k_1\leftrightarrow k_2\},
\end{eqnarray}
here we have used the same definition for PV-functions expressed in (\ref{c0},\ref{b0}) and (\ref{bmu}).
The SM case is achieved when ${\cal G}_{h WW}=1$.

Contribution for the charged Higgs boson loops (Figure \ref{h-Hc-ZZ}):

\begin{eqnarray}
{\cal F}_{H^\pm}^g&=&\frac{4 g {\cal G}_{h H^+H^-} m_W c_{2W}^2} {c^2_W}\Big\{k_s^4
+\Big[ k_s^4 (-k_s^4 k_1\cdot k_2+[k_1\cdot k_2]^3+2 k_s^4 m_{H^+}^2)\nonumber\\
&&+k_1^2 k_s^4 ([k_1\cdot k_2]^2-k_s^4)\Big]C_0(k_1,k_2)+k_s^4
(-[k_1\cdot k_2]^2-k_1^2 k_1\cdot
k_2+k_s^4)B_0(k_1,p)\Big\}\nonumber\\
&&+\{k_1\leftrightarrow k_2\},
\end{eqnarray}
for the SM case this form factor is equal to zero.

Contribution for the scalar-$Z$ boson loops (Figure \ref{h-ZZ-in}):

\begin{eqnarray}
{\cal F}_S^g&=&\left.\sum_{i=h,H}\frac{k_s^4 {\cal G}_{\phi_iWW}}{9c_W^3m_Z(k_s^4-[k_1\cdot k_2]^2)}
\right\{{\cal G}_{\phi_iWW}{\cal G}_{hWW}\Big[2k_1^2k_s^4[11(k_s^4-[k_1\cdot k_2]^2)\nonumber\\
&&+3(m_{\phi_i}^2-m_Z^2)^2]-18 k_s^4(k_s^4-[k_1\cdot k_2]^2)(m_{\phi_i}^2-k_1\cdot k_2)\Big]\nonumber\\
&&+\delta_{hi}\Big[-k_1^2k_s^4[2(k_s^4-[k_1\cdot
k_2]^2)-3(m_{\phi_i}^2-m_Z^2)^2]+9k_s^4(k_s^4-[k_1\cdot k_2]^2)(m_{\phi_i}^2+m_Z^2)\Big]\nonumber\\
&&+9C_0(k_1^2,k_2^2,p^2,m_Z^2,m_{\phi_i}^2,m_Z^2){\cal G}_{\phi_iWW}{\cal G}_{hWW}(k_s^4-[k_1\cdot k_2]^2)\Big[k_1^4[2(m_{\phi_i}^2-m_Z^2)k_1\cdot k_2\nonumber\\
&&+[k_1\cdot k_2]^2-k_s^4+(m_{\phi_i}^2-m_Z^2)^2]+2k_1^2[-2k_1\cdot k_2(k_s^4-m_{\phi_i}^4+m_{\phi_i}^2m_Z^2)\nonumber\\
&&+(4m_{\phi_i}^2-3m_Z^2)[k_1\cdot k_2]^2+2[k_1\cdot k_2]^3+m_Z^2(k_s^4+(m_{\phi_i}^2-m_Z^2)^2)]+2k_1\cdot k_2[(m_Z^3-m_{\phi_i}^2m_Z)^2\nonumber\\
&&-k_s^4(m_{\phi_i}^2-2m_Z^2)]-[k_1\cdot k_2]^2(4k_s^4-3m_{\phi_i}^4+2m_{\phi_i}^2m_Z^2+m_Z^4)+2(3m_{\phi_i}^2-2m_Z^2)[k_1\cdot k_2]^3\nonumber\\
&&+3[k_1\cdot k_2]^4+k_s^4(k_s^4-m_{\phi_i}^4+2m_{\phi_i}^2m_Z^2-5m_Z^4)\Big]+[B_0(k_1^2,m_{\phi_i}^2,m_Z^2)-B_0(p^2,m_Z^2,m_Z^2)]\nonumber\\
&&\times 3{\cal G}_{\phi_i WW}{\cal G}_{hWW}\Big[3 k_1^4 ([k_1\cdot k_2]^2-k_s^4) (k_1\cdot k_2+m_{\phi_i}^2-m_Z^2)+k_1^2 [-3 m_Z^2 ([k_1\cdot k_2]^2 - k_s^4) \nonumber\\
&&\times(k_1\cdot k_2 - 2 m_{\phi_i}^2)+6 m_Z^4 (k_s^4 - [k_1\cdot k_2]^2)+3 k_1\cdot k_2 (3 [k_1\cdot k_2]^3 + 3 [k_1\cdot k_2]^2 m_{\phi_i}^2 - 2 k_1\cdot k_2 k_s^4\nonumber\\
&&- 3 k_s^4 m_{\phi_i}^2)]-m_Z^2 ([k_1\cdot k_2]^2 - k_s^4) [3 k_1\cdot k_2 (k_1\cdot k_2 - 2 m_{\phi_i}^2) - k_s^4]+6 k_1\cdot k_2 m_Z^4 (k_s^4 - [k_1\cdot k_2]^2)\nonumber\\
&&+k_1\cdot k_2 [k_1\cdot k_2 m_{\phi_i}^2 (9 [k_1\cdot k_2]^2 - 8 k_s^4)+3 (3 [k_1\cdot k_2]^4 - 4 [k_1\cdot k_2]^2 k_s^4 + k_s^8)]\nonumber\\
&&+3 k_1\cdot k_2 k_2^2 ([k_1\cdot k_2]^3 + [k_1\cdot k_2]^2 (m_{\phi_i}^2 - m_Z^2) - 2 k_1\cdot k_2 k_s^4 +k_s^4 (m_Z^2 - m_{\phi_i}^2))\Big]\nonumber\\
&&+[B_0(k_1^2,m_{\phi_i}^2,m_Z^2)-B_0(0,m_Z^2,m_Z^2)]\times 3 k_2^2 k_s^4 m_Z^2 (m_{\phi_i}^2-m_Z^2)(2 {\cal G}_{\phi_i WW}{\cal G}_{hWW}-\delta_{hi})\nonumber\\
&&-[B_0(k_1^2,m_{\phi_i}^2,m_Z^2)-B_0(0,m_{\phi_i}^2,m_{\phi_i}^2)]\times 3 k_2^2 k_s^4 m_{\phi_i}^2 (m_{\phi_i}^2 - m_Z^2)(2 {\cal G}_{\phi_i WW}{\cal G}_{hWW}-\delta_{0h})\nonumber\\
&&+[B_0(p^2,m_Z^2,m_Z^2)-B_0(0,m_Z^2,m_Z^2)]\times 6{\cal G}_{\phi_i WW}{\cal G}_{h WW}k_s^4m_Z^2(k_s^4-[k_1\cdot k_2]^2)\nonumber\\
&&-[B_0(p^2,m_Z^2,m_Z^2)-B_0(0,m_{\phi_i}^2,m_{\phi_i}^2)]\times 6{\cal G}_{\phi_i WW}{\cal G}_{h WW}k_s^4m_{\phi_i}^2[k_1\cdot k_2]^2\nonumber\\
&&+k_s^4B_{k_1\mu_R}(m_{\phi_i})\Big[3 k_1^2 ({\cal G}_{\phi_iWW} {\cal G}_{hWW} (5 k_s^4-8 [k_1\cdot k_2]^2)+\delta_{hi} ([k_1\cdot k_2]^2-k_s^4))\nonumber\\
&&-3 ({\cal G}_{\phi_iWW} {\cal G}_{hWW} (6 [k_1\cdot k_2]^3-6 k_1\cdot k_2 k_s^4+k_s^4 m_{\phi_i}^2)+2 \delta_{hi} ([k_1\cdot k_2]^2-k_s^4) (m_{\phi_i}^2-5 m_Z^2))\nonumber
\end{eqnarray}
\begin{eqnarray}
&&+9 {\cal G}_{\phi_iWW} {\cal G}_{hWW} k_2^2 k_s^4\Big]+3 k_s^4 m_Z^2 \delta_{hi} (k_s^4-[k_1\cdot k_2]^2)B_{0\mu_R}(m_Z)\nonumber\\
&&-3 k_s^4 m_{\phi_i}^2 [k_s^4 (2 {\cal G}_{\phi_iWW} {\cal G}_{hWW}-\delta_{hi})+[k_1\cdot k_2]^2 \delta_{hi}]B_{0\mu_R}(m_{\phi_i})\nonumber\\
&&+3 {\cal G}_{\phi_iWW} {\cal G}_{hWW} k_s^4 m_{\phi_i}^2 (2
[k_1\cdot k_2]^2-3
k_s^4)B_{p\mu_R}(m_Z)\Big\}+\{k_1\leftrightarrow k_2\},
\end{eqnarray}
here we have used a new definitions for the renormalizable PV-functions
\begin{eqnarray}\label{renor2}
B_{k_1\mu}(m)&=&m_Z^2C_0(k_1^2,0,k_1^2,m^2+\mu_R^2,m_Z^2+\mu_R^2,\mu_R^2)+B_0(k_1^2,\mu_R^2,m^2+\mu^2)- B_0(0,\mu_R^2,m_Z^2+\mu_R^2)\nonumber\\
&&+\frac{m_Z^2+\mu^2}{m_Z^2}\Big(B_0(0,m_Z^2+\mu_R^2,m_Z^2+\mu_R^2)-B_0(0,\mu_R^2,\mu_R^2)\Big)+1,
\end{eqnarray}
\begin{eqnarray}\label{renor3}
  B_{p\mu_R}(m)&=&B_0(p^2,m^2+\mu_R^2,m^2+\mu_R^2)-B_0(0,\mu_R^2,\mu_R^2),\\
  B_{0\mu_R}(m)&=&B_0(0,m^2+\mu_R^2,m^2+\mu_R^2)-B_0(0,\mu_R^2,\mu_R^2).
\end{eqnarray}
The SM context is attained when $i=h$.

Contribution for the triple Higgs boson loops (Figure \ref{hhh-ZZ}):
\begin{eqnarray}
{\cal F}_{3\phi}^g&=&\sum_{i,j=h,H}\left.\frac{3k_s^4 {\cal
G}_{h\phi_i\phi_j}}{2c_W^2 m_W} \right\{{\cal
G}_{\phi_iWW}{\cal G}_{\phi_jWW}\Big[2k_s^4+
\Big\{k_1^2\Big(-2(m_{\phi_j}^2-m_Z^2)k_1\cdot k_2+[k_1\cdot k_2]^2-k_s^4\nonumber\\
&&+(m_{\phi_j}^2-m_Z^2)^2\Big)+k_2^2\Big(-2(m_{\phi_i}^2-m_Z^2)k_1\cdot k_2+[k_1\cdot k_2]^2-k_s^4+(m_{\phi_i}^2-m_Z^2)^2\Big)\nonumber\\
&&+2\Big[-k_1\cdot k_2\Big(k_s^4+(m_{\phi_i}^2-m_Z^2)(m_Z^2-m_{\phi_j}^2)\Big)-[k_1\cdot k_2]^2(m_{\phi_i}^2+m_{\phi_j}^2-2m_Z^2)+[k_1\cdot k_2]^3\nonumber\\
&&+k_s^4(m_{\phi_i}^2+m_{\phi_j}^2-4m_Z^2)\Big]\Big\}C_0(k_1^2,k_2^2,p^2,m_{\phi_i}^2,m_Z^2,m_{\phi_j}^2)+[B_0(k_1^2,m_{\phi_i}^2 ,m_Z^2)\nonumber\\
&&-B_0(p^2,m_{\phi_i}^2,m_{\phi_j}^2)]\Big[k_1^2(m_{\phi_j}^2-m_Z^2-k_1\cdot k_2)-k_1\cdot k_2(k_1\cdot k_2-m_{\phi_i}^2+m_Z^2)\Big]\nonumber\\
&&+[B_0(k_1^2,m_{\phi_j}^2
,m_Z^2)-B_0(p^2,m_{\phi_i}^2,m_{\phi_j}^2)] \Big[k_1^2(m_{\phi_i}^2-m_Z^2-k_1\cdot k_2)\nonumber\\
&&-k_1\cdot k_2(k_1\cdot
k_2-m_{\phi_j}^2+m_Z^2)\Big]\Big]+k_s^4\Big[{\cal
G}_{\phi_iWW}{\cal
G}_{\phi_jWW}(B_{1\mu_R}(m_{\phi_i})+B_{1\mu_R}(m_{\phi_j}))\nonumber\\
&&-2\delta_{ij}B_{p\mu_R}(m_{\phi_i})\Big]\Big\}+\{k_1\leftrightarrow
k_2\},
\end{eqnarray}
the renormalized PV-function have been expressed as Eqs. (\ref{renor2}) and (\ref{renor3}).

Finally the general tree level decays for the Higgs boson needed for compute the of $R$ factor are given in detail in Ref. \cite{decay}.


\begin{thebibliography}{99}

\bibitem{Atlas}
The ATLAS Collaboration, Phys. Lett. B716, 1 (2012) and
ATLAS-CONF-2012-162; The CMS Collaboration, Phys. Lett. B716,
  30 (2012) and CMS-PAS-HIG-12-045.

\bibitem{PHiggs}
P. Higgs, Phys. Lett. 12, 132 (1964); Phys. Rev. Lett. 13,
     508 (1964); F. Englert and R. Brout, Phys. Rev. Lett. 13,
    321 (1964); G. Guralnik, C. Hagen and T. Kibble, Phys. Rev.
     Lett. 13, 585 (1964); S. Weinberg, Phys. Rev Lett. 19, 1264
   (1967).

\bibitem{Aad}
G.Aad et al. (ATLAS Collaboration, CMS Collaboration),
        Phys. Rev. Lett. 114, 191803 (2015); Phys. Rev. D91, 012006 (2015); S. Chatrchyar et al.(CMS Collaboration), JHEP 1401, 096 (2014);
         M.~Flechl [ATLAS and CMS Collaborations],
  J.\ Phys.\ Conf.\ Ser.\  {\bf 631}, no. 1, 012028 (2015)
  [arXiv:1503.00632 [hep-ex]].


\bibitem{Aguilar}
 J.A. Aguilar Saavedra et al., ECFA-DESY LC Physics Working
      Group Collaboration. hep-ph/0106315.

\bibitem{Gutierrez}
A. Gutierrez-Rodriguez, M.A. Hernandez-Ruiz, O.A. Sampayo,
     Int. J. Mod. Phys. A24, 5299 (2009).

\bibitem{Baur}
U. Baur, T. Plehn and D.L. Rainwater, Phys. Rev. D67, 112 (2003);
     J.~Baglio, A.~Djouadi, R.~Gröber, M.~M.~Mühlleitner, J.~Quevillon and M.~Spira,
    JHEP {\bf 1304} (2013) 151
  [arXiv:1212.5581 [hep-ph]].

\bibitem{Djouadi}
A. Djouadi et al., Eur. Phys. J. C10, 45 (1999); S. Dawson, S.
   Dittmaier and  M. Spira, Phys. Rev. D58, 115012 (1998).

\bibitem{Dolan}
 M.J. Dolan, C. Englert, and M. Spannoswsky, JHEP 1210, 112 (2012).

\bibitem{Baglio}
J.~Baglio, A.~Djouadi, R.~Gröber, M.~M.~Mühlleitner, J.~Quevillon
and M.~Spira, JHEP {\bf 1304} (2013) 151
  [arXiv:1212.5581 [hep-ph]]; J.~Baglio,
  Pos DIS {\bf 2014}, 120 (2014)
  [arXiv:1407.1045 [hep-ph]].

\bibitem{Dawson}
     S.~Dawson, A.~Ismail and I.~Low,
  Phys.\ Rev.\ D {\bf 91}, no. 11, 115008 (2015)
  [arXiv:1504.05596 [hep-ph]].

\bibitem{Maltoni}
 F. Maltoni, E. Vryonidou, M. Zaro, arXiv:1408.6542,
          JHEP 1411, 079 (2014).

\bibitem{He}
 H.J. He, J. Ren and W. Yao, arXiv:1506.03302.

\bibitem{Goertz}
  F. Goertz et al., arXiv:1301.3492, JHEP 1306, 016 (2013).

\bibitem{Flechl}
 M.~Flechl [ATLAS and CMS Collaborations],
  J.\ Phys.\ Conf.\ Ser.\  {\bf 631}, no. 1, 012028 (2015)
  [arXiv:1503.00632 [hep-ex]].

\bibitem{Martinez}
R. Martinez, M.A. Perez, J.J. Toscano,
        Phys.Lett. B340, 91 (1994); F. Larios,
                  R. Martinez, M.A. Perez, Phys. Lett. B345,
                  259 (1995)

\bibitem{Kanemura1}
  S.~Kanemura, M.~Kikuchi and K.~Yagyu,
  Nucl.\ Phys.\ B {\bf 896}, 80 (2015)
  [arXiv:1502.07716 [hep-ph]]. S. Kanemura et al., Phys. Lett. B558, 157 (2003)

\bibitem{Heinemeyer}
S.~Heinemeyer {\it et al.} [LHC Higgs Cross Section Working Group Collaboration],
  arXiv:1307.1347 [hep-ph].

\bibitem{Cao}
 Q.~H.~Cao, Y.~Liu and B.~Yan,
  arXiv:1511.03311 [hep-ph].


\bibitem{Arhrib}
   A.~Arhrib, R.~Benbrik, J.~El Falaki and A.~Jueid,
  JHEP {\bf 1512}, 007 (2015)
  [arXiv:1507.03630 [hep-ph]].



\bibitem{Castilla}      H. Castilla-Valdez, C.G. Honorato, A. Moyotl,
        M.A. Perez (in preparation).

\bibitem{Figy}      V.~Hankele, G.~Klamke, D.~Zeppenfeld and T.~Figy,
  Phys.\ Rev.\ D {\bf 74}, 095001 (2006)
  [hep-ph/0609075].



\bibitem{Bager}
Vernon D. Bager, Roger J. N. Phillips {\it Collider
Physics},(Addison-Wesley, 1997).



\bibitem{Gunion}
J.F. Gunion and H.E. Haber, Phys. Rev. D67, 075019 (2003).

\bibitem{HernandezSanchez}
  J.~Hernandez-Sanchez, C.~G.~Honorato, M.~A.~Perez and J.~J.~Toscano,
  Phys.\ Rev.\ D {\bf 85}, 015020 (2012)
  [arXiv:1108.4074 [hep-ph]].



\bibitem{Cordero-Cid}
   A.~Cordero-Cid, J.~Hernandez-Sanchez, C.~G.~Honorato, S.~Moretti, M.~A.~Perez and A.~Rosado,
  JHEP {\bf 1407}, 057 (2014)
  [arXiv:1312.5614 [hep-ph]].

\bibitem{Chatrchyan}
  S.~Chatrchyan {\it et al.}  [CMS Collaboration],
  arXiv:1312.5353 [hep-ex].


\bibitem{HernandezSanchez:2012eg}
  J.~Hernandez-Sanchez, S.~Moretti, R.~Noriega-Papaqui and A.~Rosado,
  JHEP {\bf 1307}, 044 (2013)
  doi:10.1007/JHEP07(2013)044
  [arXiv:1212.6818].

\bibitem{Pittau:2012zd}
  R.~Pittau,
  JHEP {\bf 1211}, 151 (2012)
  [arXiv:1208.5457 [hep-ph]].


\bibitem{decay}
J.F. Gunion, H.E. Haber, G.L. Kane and S. Dawson, The Higgs Hunter's Guide,
Addison-Wesley, Reading, MA (1990).  W.-Y. Keung and W.J. Marciano, Phys. Rev. D 30
(1984).  E. Barradas, J.L. Diaz-Cruz, A. Gutierrez and A. Rosado,  Phys. Rev. D 53 (1996) 1678.
J.L. Diaz-Cruz and M.A. Perez, Phys. Rev. D 33
(1986) 273.  M. Gomez-Bock and R. Noriega-Papaqui, J. Phys. G 32 (2006) 761 [hep-ph/0509353].


\end{thebibliography}
\end{document}